\documentclass{JHEP3}
\usepackage{amsmath}
\usepackage{amsthm}
\usepackage{amsfonts}
\usepackage{amssymb}
\def\KZ{{\text{H}}}
\def\BPZ{{\text{L}}}
\def\KZB{{\text{H}}}

\def \Fig#1#2#3 {
\begin{figure}
\centering
\epsfxsize=#2cm \epsfbox{#1.eps}
\caption{#3}
\label{#1}
\end{figure}
}

\newcount\figno
\def\fig#1#2#3{
\par\begingroup\parindent=0pt\leftskip=1cm\rightskip=1cm\parindent=0pt
\baselineskip=15pt
\global\advance\figno by 1
\epsfxsize=#3
\centerline{\epsfbox{#2}}
\vskip 12pt
{\bf \small Figure \the\figno:} {\small #1}\par
\endgroup\par
}
\def\figlabel#1{\xdef#1{\the\figno
\mbox{ }}}
\def\encadremath#1{\vbox{\hrule\hbox{\vrule\kern8pt\vbox{\kern8pt
\hbox{$\displaystyle #1$}\kern8pt}
\kern8pt\vrule}\hrule}}

\def\b{{\beta}}
\def\bb{\bar{\b}}
\def\c{{\gamma}}
\def\bc{\bar{\c}}
\def\pl{\partial}
\def\bpl{\bar{\pl}}
\def\bz{\bar{z}}

\def\brangle{\right\rangle}
\def\blangle{\left\langle}
\def\vardelta{\boldsymbol{\delta}}

\title{$H^+_3$ WZNW model from Liouville field theory}
\author{Yasuaki Hikida and Volker Schomerus\\
DESY Theory Group, DESY Hamburg, Notkestrasse 85, D-22603 Hamburg, Germany\\
E-mail: \email{yasuaki.hikida@desy.de}, \email{volker.schomerus@desy.de}
}


\abstract{
There exists an intriguing relation between genus zero
correlation functions in the $H^+_3$ WZNW model and in Liouville
field theory. We provide a path integral
derivation of the correspondence and then use our new approach to
generalize the relation to surfaces of arbitrary genus $g$. In
particular we determine the correlation functions of $N$ primary
fields in the WZNW model explicitly through Liouville correlators
with $N + 2g-2$ additional insertions of certain degenerate
fields. The paper concludes with a list of interesting further
extensions and a few comments on the relation to the geometric
Langlands  program.}

\preprint{DESY 07-075}

\begin{document}



\section{Introduction}

The $H^+_3$ Wess-Zumino-Novikov-Witten (WZNW) model has received
considerable attention as an interesting non-rational conformal
field theory \cite{Teschner1,Teschner2} and as the Euclidean
version of $AdS_3$ \cite{OM3}. The 3-dimensional target space of
the theory may be parametrized by coordinates $\gamma$ and $\bar
\gamma$ of its 2-dimensional boundary along with some radial
coordinate $\phi$. As for any Anti-de Sitter (AdS) geometry, the
latter is particularly interesting. Physically, it should be
regarded as a very close relative of the Liouville direction in
2-dimensional string theory.%
\smallskip

It was long observed \cite{Teschner1} that the $3$-point functions
of the $H^+_3$ WZNW model coincide with those of Liouville field
theory \cite{DoOt,ZaZa,Teschner:1995yf} up to some simple kinematical 
factors that
are determined by the SL(2,${\mathbb C}$) symmetry of the $H^+_3$ background.
In particular, all the highly nontrivial (non-perturbative)
curvature dependence of the $H^+_3$ $3$-point couplings is
inherited from the stringy corrections of Liouville theory. A very
remarkable generalization of this fact was discovered in
\cite{RT}: Ribault and Teschner determined genus zero correlators
for any number $N$ of WZNW primaries in terms of certain $2N-2$
point functions in Liouville theory. Their proof relies on a
relation \cite{Stoyanovsky} between differential equations of both
theories, namely the Knizhnik-Zamolodchikov (KZ) equations for
WZNW models and the Belavin-Polyakov-Zamolodchikov (BPZ) equations
for Liouville correlators, along with descent relations provided
by the explicit knowledge of the $3$-point functions. The relation
between Liouville theory and the $H^+_3$ model was further explored 
in \cite{Giribet:2005xr,Giribet:2005ix,Nakamura:2005sm}. 
\smallskip

The aim of this paper is twofold. To begin with, we shall
re-derive the Ribault-Teschner correspondence for genus zero
correlation functions using a rather elegant path integral
computation instead of heavy algebraic manipulations. Our
derivation provides a completely new view on the map between WZNW
and Liouville primaries and on the necessity to introduce $N-2$
further degenerate fields in the Liouville correlation function.
The simplicity of our derivation opens the way to various
generalizations. Among them is the extension of the correspondence
to correlators on higher genus surfaces. In fact, using
essentially the same ideas as for the tree-level computation, we
shall derive a precise expression for $N$-point correlators of
WZNW primaries on any closed Riemann surface in terms of Liouville
field theory. On the Liouville side, the construction involves
correlation functions with $N+2g-2$ additional insertions of
degenerate fields. Further extensions, e.g.\ to the case with
world-sheets with boundaries or target space groups of rank $r >
1$ will be briefly discussed at the end of this work.
\smallskip

Let us now describe the contents of each section and our main
results in a bit more detail. As we have mentioned, we shall start
our analysis by giving a new derivation of the Ribault-Teschner
correspondence on the sphere. For the convenience of the reader,
we also include a few comments on the relation between
differential equations. Section 3 contains the generalization to
the torus. For genus $g=1$, all the special functions involved in
the formulation of the correspondence can be easily expressed in
terms of Jacobi's $\theta$ function. This makes our formulas
particularly easy to deal with. In particular, we shall be able to
demonstrate explicitly how our correspondence intertwines between
the relevant differential equations. Correlation functions of
Liouville theory on the torus satisfy an extension of the BPZ
differential equation \cite{EO} that involves a derivative with
respect to the modular parameter $\tau$ of the torus in addition
to derivatives with respect to the insertion points of fields. On
the WZNW side, the story is a bit more complicated. In fact, the
usual WZNW correlators do not obey KZ-type equations. The problem
arises from the zero modes of currents which cannot be written as
differential operators acting on correlation functions. This issue
was resolved by Bernard in \cite{Bernald}, who had the idea to
introduce an additional dependence on the choice of some group
element. The latter parametrizes possible twisted boundary
conditions for currents along the $\beta$-cycle of the torus. Not
all these {\it dim} G parameters are actually needed for the
formulation of KZ-type equations on the torus. The minimal number
of required extra parameters is given by the rank of $G$ rather
than its dimension. In our case this means that one extra twist
parameter $\lambda$ is sufficient. Once such an extension of WZNW
correlators is taken into account, they satisfy the
Knizhnik-Zamolodchikov-Bernard (KZB) differential equations. By
our correspondence, these are correctly related to the genus $g=1$
BPZ equations.
\smallskip

Section 4 contains the main new formula of this work, namely a
construction of $N$-point WZNW correlators on a closed Riemann
surface of genus $g$ from correlation functions in Liouville field
theory, see eq.\ (\ref{corrga}). As we stated before, the latter
involves $N+2g-2$ insertions of Liouville degenerate fields in
addition to the $N$ primary fields. It is instructive to compare
the parameter spaces on both sides of this correspondence.
Obviously, the correlation functions of the WZNW model must be
extended by introducing extra parameters generalizing the role of
the modulus $\lambda$ we discussed at length in the previous
paragraph. It turns out that on a surface of genus $g \geq 1$ we
need $2g-1$ new complex coordinates. For fixed surface moduli
$\tau$ and insertion points $z_\nu, \nu=1,\dots,N$, an $N$-point
function of WZNW primaries therefore depends on real $3N$ target
space momenta $(j_\nu, \mu_\nu,\bar \mu_\nu)_{\nu=1, \dots,N}$ and
$2g-1$ complex moduli. These add up to $3N+4g-2$ real moduli. On
the Liouville side, we count $N$ real target space momenta
$\alpha_\nu$ in addition to the (complex) position of $N+2g-2$
Liouville degenerate fields and an overall complex pre-factor $u$.
The total number of real parameters is therefore $3N+4g-2$, just
as in the WZNW theory.

\section{Derivation of the correspondence - genus 0}

Our first task is to explain how the correspondence emerges from a
path integral `definition' of the WZNW model. Indeed, we will be
able to recover the formula of Ribault and Teschner through some
formal path integral manipulations. In the second subsection we
briefly review how the differential equations on both sides of the
correspondence are mapped onto each other. At genus 0 this is not
new, but it will help to understand the corresponding analysis at
higher genus later on. Finally, we shall comment on the hidden
problems of the formal path integral approach and the precise
interpretation of our results.

\subsection{Path integral derivation of the correspondence}

Let us begin with a little bit of background on the $H^+_3$ WZNW
model. In the most common presentation, the action involves the
three fields $\gamma, \bar \gamma$ and $\phi$ corresponding to the
coordinates that parametrize the 2D boundary of $H^+_3$ and the
radial direction, respectively. We will not work with this version
but pass to a first order formulation which
includes two additional fields $\beta$ and $\bar \beta$ of
conformal weight $(h_\b,0) = (1,0)$ and $(0,h_{\bb}) = (0,1)$.
Throughout this work we shall work in conformal gauge, where the
world-sheet metric and curvature are determined by some function
$\rho$ through
$$ ds^2 \ = \ |\rho(z)|^2 \, dz d\bar z \ \ \ \ , \ \ \
\sqrt g {\cal R} \ = \ - 4 \partial \bar \partial \ln |\rho|^2\ \
.
$$
In this particular gauge, the action of the $H^+_3$ WZNW model
takes the following form (see e.g.\ 
\cite{Giveon:1998ns,Hosomichi:2000bm})
\begin{equation} \label{action}
S[\phi,\c,\b] \ = \ \frac{1}{2\pi} \int d^2w \, \left(
\bpl \phi \pl \phi -  \b \bpl \c - \bb \pl \bc +
\frac{Q_\phi}{4} \, \sqrt{g} {\cal R}\, \phi - b^2 \b\bb
e^{2b\phi}\right)
\ \ .
\end{equation}
Here, the parameter $b$ is related to the level of the WZNW model
through $b^{-2} = k-2$ and the background charge $Q_\phi$ is given
by $Q_\phi=b$. Let us note that the usual action of the $H^+_3$
model emerges after integration over $\b$ and $\bb$. The total
central charge of the model is computed from the level $k$ using
$$ c(H^+_3) \ = \ \frac{3k}{k-2} \ = \ 3 + 6b^2 \ = \
  2 + (1+6b^2)\ \ .$$
In the last step we split the central charge into the contribution
$c(\b\c) =2$ from the $\b\c$ system and the remainder $c(\phi) = 1
+ 6 b^2$ which originates from the bosonic field $\phi$ with
background charge $Q_\phi = b$.
\smallskip

The second ingredients we shall need before we can study correlators
of the $H^+_3$ model are the vertex operators of the theory. In the
so-called $\mu$-basis these read
\begin{align} \label{vertex}
 V_j ( \mu | z ) \equiv |\rho(z)|^{2\Delta^H_j}\, | \mu |^{ 2 j + 2}
 \, e^{ \mu \gamma(z) - \bar \mu \bar \gamma(\bar z)} e^{2b (j +1 )
\phi(z,\bz)} ~.
\end{align}
In conformal gauge, the factor involving $\rho(z)$ must be included
in order for $V$ to transform as a primary of weight zero under
conformal transformations. The quantity $\Delta^H_j$ in the
exponent is given by
$$ \Delta^H_j \ = \ -b^2 j(j+1) \ \ . $$
One may consider $\mu$ and $\bar \mu$ as Euclidean light cone
momenta. Similarly, the parameter $j$ is related to the momentum
in radial $\phi$ direction. In the $\mu$ basis, the operator
products of currents with primary fields are given by
\begin{align} \label{opejv}
J^a(w) \  V_j ( \mu | z ) \ = \ (w-z)^{-1} \, D^a  V_j ( \mu | z ) \ \ \
   \mbox{ for } \ \ a \ = \ \pm,0
\end{align}
where the generators $D^a$ of global target space symmetries
take the form
\begin{equation}
 D^- = \mu ~, \qquad
D^0 = - \mu \partial_\mu ~,  \qquad D^+ = \mu \partial^2_\mu -
\frac{j(j+1)}{\mu}  \label{mubasis}
\end{equation}
and similarly for the remaining three generators $\bar D^a$. 
The $\mu$ basis is the most convenient one for what we
are about to discuss.
\smallskip

Our aim is to compute the $N$-point function of primary fields in
the $H^+_3$ model, i.e.\
$$ \blangle \ \prod_{\nu=1}^{N} \, V_{j_\nu}(\mu_\nu|z_\nu) \ \brangle^H
\ = \ \int {\cal{D}} \phi {\cal{D}} \c {\cal{D}} \b \
e^{-S[\phi,\c,\b]} \prod_{\nu=1}^{N} \, V_{j_\nu}(\mu_\nu |z_\nu)
\ \ . $$ Our first step is to integrate out the fields $\c$ and
$\bc$. This is rather easy because they appear only linearly in
both the action and the exponents of the vertex operators. Hence,
the integration leads to a simple $\delta$ function constraint on
the coefficients of $\c$ and $\bc$, i.e.\
\begin{equation} \label{plb}
\bpl \b(w) \ = \ 2\pi \sum_{\nu=1}^N \mu_\nu \delta^2(w-z_\nu)
\ \ \ \ , \ \ \ \ \pl \bar\b(\bar w) \ = \
  -  2\pi \sum_{\nu=1}^N \bar \mu_\nu \delta^2(\bar w- z_\nu) ~.
\end{equation}
The distribution $\delta^2$ on the right hand side is normalized such
that $\int d^2 z \delta^2 (z) f(z,\bar z) \ = \ f(0)$. Let us stress
that a meromorphic differential $\beta$ with the property
(\ref{plb}) exists if and only if the sum $\sum \mu_\nu$ vanishes.
Once this condition is met, the integration of $\bpl \beta$ and
$\pl \bb$ with respect to the world-sheet coordinate $w$ is
immediately performed using the simple auxiliary formulas
$$ \bpl (1/z) \ = \ \pl (1/\bar z) \ = \ 2 \pi
\delta^2(z) \ \ .
$$
The result is
\begin{equation}
\b(w) \ = \  \sum_{\nu=1}^N \frac{
\mu_\nu}{w-z_\nu} \ \ \ .
\end{equation}
A similar equation holds for $-\bb$. The crucial idea
now is to re-parametrize $\beta$ using simple facts about
meromorphic one-differentials on the sphere. To begin with, we recall
that for any one-differential, the number of poles exceeds the number
of zeroes by two. Hence $\beta$ must have $N-2$ zeroes whose
locations on the sphere we denote by $w=y_i$. Furthermore, a
differential is uniquely characterized by the position of its
zeroes and poles up to an overall factor $u$. Consequently, we can
rewrite $\b$ in the form
\begin{equation} \label{bspb}
\b(w) \ = \ \ u \, \frac{\prod_{i=1}^{N-2}
(w-y_i)}{\prod_{\nu=1}^N (w-z_\nu)} \ = : \ u \,
\mathcal{B}_0 (y_i,z_\nu;w) \ \ .
\end{equation}
Thereby, we have now replaced the $N$ parameters $\mu_\nu$ subject
to constraint $\sum_\nu \mu_\nu =0$ through $N-2$ coordinates
$y_i$ and a global factor $u$. We can recover the residues
$\mu_\nu$ of $\beta$ from the new variables $y_i$
and $u$ through
\begin{equation} \label{mufromuy}
\mu_\nu \ =\  u \, \frac{\prod_{j=1}^{N-2}
( z_\nu - y_j)} {\prod_{\mu \neq \nu}^N (z_\nu - z_\mu)} 
\ \ .
\end{equation}
The new variables may be used to rewrite the $\vardelta$ function
resulting from the integration over $\gamma$ and $\bar\gamma$,
\begin{equation} \label{delta0}
 \vardelta^2 (\bar \partial \beta (w) -
    2\pi \sum_{\nu=1}^N \mu_\nu \delta^2(w-z_\nu) ) \ = \
     \delta^2(\sum_{\nu=1}^N \mu_\nu) \
   \vardelta^2\left(\beta - u {\cal B}_0(y_i,z_\mu,w)\right)\ \ .
\end{equation}
The Jacobian for the transformation from
$\bar\partial\beta,\pl\bar\beta$ to $\beta,\bar \beta$ is trivial
on the sphere. Once eq.\ (\ref{delta0}) has been inserted into our
path integral, we perform the integral over the fields $\b$ and
$\bar \b$ to obtain
\begin{eqnarray*}
 \blangle \ \prod_{\nu=1}^{N} \, V_{j_\nu}(\mu_\nu|z_\nu) \ \brangle^H
& = & |u|^2 \, \delta^2(\sum_{\nu=1}^N \mu_\nu) \int {\cal{D}}
\phi e^{-\frac{1}{2\pi} \int d^2w \  \left( \bpl \phi \pl \phi +
\frac{Q_\phi}{4} \sqrt{g} {\cal R} \phi +  b^2
|{\mathcal{B}_0}|^2\, e^{2 b \phi}\right) }\,
\times \\[2mm] & & \hspace*{0cm} \times
\prod_{\nu=1}^{N} |\rho(z_\nu)|^{2\Delta^H_\nu} \, |u|^{-2(j_\nu+1)}
|\mu_\nu|^{ 2(j_\nu + 1)} e^{2b (j_\nu +1 ) \phi(z_\nu)}\ .
\end{eqnarray*}
In writing this formula we have also shifted the zero mode of the
bosonic field $\phi$ by $\phi \mapsto \phi - (1/b)\ln |u|$. This
removes the $u$ dependence from the interaction term but introduces
an additional factor $|u|^2$ through the coupling of $\phi$ to the
world-sheet curvature.%
\smallskip

In order to prepare for the second step of our calculation we
observe that the exponential field $\exp (2b\phi)$ always comes
multiplied with $|{\mathcal{B}_0}|^2$. For the corresponding term
in the action, this has been spelled out explicitly. In the case
of the vertex operators, the prefactor is hidden in $|\mu_\nu|^2$
once we insert eq.\ (\ref{mufromuy}). This suggests to introduce a
new bosonic field $\varphi$ through
\begin{equation}
\varphi(w,\bar w) \ := \ \phi(w,\bar w) + \frac{1}{2b} \left(
\sum_i \ln |w-y_i|^2 - \sum_\nu \ln |w-z_\nu|^2 - \ln |\rho(w)|^2
\right) \ \
\end{equation}
where the term in brackets is $\ln |\mathcal{B}_0|^2$. Acting with
$\pl \bpl$ gives
$$
\pl \bpl \varphi(w,\bar w) \ = \ \pl \bpl \phi(w,\bar w) +
\frac{\pi}{b} \left(\sum_i \delta^2(w-y_i) - \sum_\nu
\delta^2(w-z_\nu) \right) - \frac{1}{2b}\partial\bar \partial \ln
|\rho(w)|^2 \ \ .
$$
Before we insert our formulas for $\varphi$ we have to comment on
one small subtlety. In fact, at various places we will have to
evaluate the quantity $\ln|w-z_\nu|$ at the point $w=z_\nu$. This
leads to well-known singularities which need to be regularized.
Following Polyakov \cite{Polyakov}, we shall use the prescription
\begin{equation} \label{reg}
   \left( \lim_{w \rightarrow z} \ln|w-z|^2\right)_{\text{fin}}
   \ := \ - \ln |\rho(z)|^2 \ \ .
\end{equation}
The right hand side is finite and has the same behavior under
conformal transformations as the divergent left hand side.
Using our change of variables from $\phi$ to $\varphi$ along
with the rule (\ref{reg}) it is easy to see that
\begin{equation} \label{regexpz}
  |\mu_\nu|^{ 2(j_\nu + 1)} e^{2b (j_\nu +1 ) \phi(z_\nu)}
 \ = \
 |u|^{2(j_\nu+1)}  \ |\rho(z_\nu)|^{-2(j_\nu+1)} \
 e^{2 b (j_\nu+1) \varphi(z_\nu)}\ \ .
\end{equation}
After these comments we are prepared to perform our
substitution from $\phi$ to $\varphi$. We do not want to
spell out all the details, but let us observe that
$$ -\frac{1}{2\pi} \int d^2 w\, \bar \partial \phi \partial
\phi \ = \ - \frac{1}{2\pi} \int  d^2 w\,\left(  \bar \partial
 \varphi  \partial \varphi + \frac{1}{4b} \sqrt{g}{\cal R} \varphi
  \right) - \frac{1}{b} \left( \sum_{i=1}^{N-2} \varphi(y_i)
  - \sum_{\nu=1}^N \varphi(z_\nu)\right) + \dots
$$
where the $\dots$ stand for terms that do not contain the
field $\varphi$. There are a few things we can read off from
this result. To begin with, the background charge $Q_\phi$ of
the bosonic field $\phi$ gets shifted by an amount $\Delta Q =
1/b$ to the new background charge $Q_\varphi = b + 1/b$.
Furthermore, the exponents of the vertex operator insertions
at $z_\nu$ are all shifted by $\varphi(z_\nu)/b$. The
precise relation is
$$ e^{2b(j_\nu+1)\varphi(z_\nu)} \ e^{\frac{1}{b} \varphi(z_\nu)}
\ = \  |\rho(z_\nu)|^{2(j_\nu+1)} \,  e^{2b(j_\nu+1 +
 \frac{1}{2b^2})\varphi(z_\nu)}  \ \ .
$$
Note that the additional power of $\rho$ is required to
match the behavior of both sides under conformal
transformations. More strikingly, new vertex operators
$\exp (-\varphi(w)/b)$ are inserted at the $N-2$ points
$y_i$. With a little bit of additional care we can also
work out the additional $\varphi$-independent factors.
The result is
\begin{eqnarray*}
\blangle \ \prod_{\nu=1}^{N} \, V_{j_\nu}(\mu_\nu|z_\nu) \ \brangle^H
& = & |\Theta_N(u,y_i,z_\nu)|^2  \, \delta^2(\sum_{\nu=1}^N
\mu_\nu)  \int {\cal{D}} \varphi \, e^{-\frac{1}{2\pi}
\int d^2w \, \left( \bpl \varphi \pl \varphi + \frac{Q_\varphi}{4}
\sqrt{g} {\cal R} \varphi + b^2 |\rho|^2 e^{2 b \varphi}\right) }\\[2mm]
& & \times\  \prod_{\nu=1}^{N} \, |\rho(z_\nu)|^{2\Delta_\nu^L}
\, e^{2\left( b (j_\nu +1) +
\frac{1}{2b}\right)  \varphi(z_\nu)} \, \prod_{i=1}^{N-2} \,
 |\rho(y_i)|^{2\Delta^L_{-1/2b}}\, e^{-\frac{1}{b} \varphi(y_i)}\ \
\end{eqnarray*}
where, following \cite{RT}, we collected various factors in the
function $\Theta_N$ defined by
\begin{align}
 \Theta_N (u,y_i,z_\nu)
  \ = \ u  \,
     \prod_{\mu < \nu}^N ( z_\mu - z_\nu )^{\frac{1}{2 b^2}}
      \prod_{i < j}^{N-2} (y_i - y_j )^{\frac{1}{2 b^2}}
      \prod_{\mu=1}^N \prod_{i=1}^{N-2}
       (z_\mu - y_i)^{- \frac{1}{2 b^2}} ~,
\end{align}
and we introduced the quantities \begin{equation} \label{ajrel}
\Delta^L_\alpha\ =\ \alpha (Q_\varphi - \alpha) \ \ \ \mbox{ for }
\ \ \ \alpha \ =\ -\frac{1}{2b}  \ \ \mbox{ or } \ \ \alpha \ = \
\alpha_\nu
  \ = \ b (j_\nu + 1) + \frac{1}{2b} \ \ . \end{equation}
Except for the first two factors, the result may now be re-expressed
in terms of correlation functions of Liouville theory,
\begin{align}
\blangle \, \prod_{\nu=1}^{N}  V_{j_\nu}(\mu_\nu|z_\nu) \,
   \brangle^{H} \, = \, \delta^2 (\sum_{\nu=1}^N \mu_\nu)
|\Theta_N(u,y_i,z_\nu)|^2 \, \blangle \prod_{\nu=1}^N
V_{\alpha_\nu} (z_\nu) \, \prod_{i=1}^{N-2}
V_{-1/2b}(y_i)\brangle^{L}.
\label{corrg0}
\end{align}
The Liouville correlator on the right hand side is evaluated with
a bulk cosmological constant $\mu_B = 4b^2$ and the primaries are
defined by \begin{equation} \label{vertexL}  V_\alpha(z) \ =\
|\rho(z)|^{2\Delta^L_\alpha} \,
     e^ {2 \alpha \varphi}\ \ .  \end{equation}
Up to an overall constant, our result coincides with the
formula found by Ribault and Teschner in \cite{RT}.

\subsection{From KZ to BPZ differential equations}

The correlation functions on both sides of the correspondence are
known to satisfy certain differential equations. Throughout this
subsection we shall set $\rho=1$. Liouville correlators are known
to obey the Belavin-Polyakov-Zamolodchikov (BPZ) second order
differential equations that come with the $N-2$ insertions
of the degenerate field $V_{-1/2b}$. In fact, this field is well
known to possess a singular vector on the second level, or,
equivalently, to satisfy the differential equation
$$ \partial_y^2 V_{-1/2b}(y) + b^{-2} :T(y) V_{-1/2b}(y):
\ = \ 0 \ \ . $$
Such an equation holds for each of the $N-2$ degenerate fields
at the points $y_i$. Using the Ward identities for the Virasoro
field $T(y_i)$, we can convert these singular vector equations
into $N-2$ second order differential equations for the Liouville
correlators $\Omega^L$ ,
\begin{equation}
D^{\BPZ}_i \, \Omega^L(z_\nu,y_i) \ = \  0  \ \ \ \
 \mbox{ where }\ \ \Omega^L(z_\nu,y_i) \ = \
\blangle \prod_{\nu=1}^N \, V_{\alpha_\nu} (z_\nu) \
\prod_{i=1}^{N-2} \, V_{-1/2b}(y_i)\brangle^{L}
\end{equation}
and the differential operators $D^{\BPZ}_i$ were first
found by Belavin, Polyakov and Zamolodchikov to be of the
form
\begin{equation}
D^{\BPZ}_i \ = \ b^2 \frac{\partial^2}{\partial y_i^2} +
\sum_{\nu=1}^N \left(\frac{\Delta^L_\nu}{(y_i - z_\nu)^2}
  + \frac{1}{y_i - z_\nu}\, \pl_\nu \right)
 +   \sum_{j\neq i}^N \left( \frac{\Delta^L_{-1/2b}}{(y_i - y_j)^2}
  + \frac{1}{y_i - y_j}\, \pl_j \right)\ .
\end{equation}
Here, $\partial_i = \partial/\partial y_i$ and $\partial_\nu
= \partial/\partial z_\nu$ and the conformal dimensions
are given by $\Delta^L_{\nu} = \alpha_\nu ( Q - \alpha_\nu )$,
as before.
\smallskip

Let us now turn to the WZNW model. Its correlators are well known
to satisfy the Knizhnik-Zamolodchikov (KZ) differential equations.
These emerge from insertions of the Sugawara singular vector
$$
T(w) + b^2 \left(\ : J^0(w) J^0(w): - \frac12(\ :J^-(w) J^+(w):+
:J^+(w) J^-(w):\ )\  \right)  \ = \ 0
$$
into WZNW correlation functions. Since we can insert this at any
point on the sphere, we certainly get an infinite number of
equations. But these are not all independent. In fact, evaluation
of the residues of the first order poles at $w=z_\nu$ gives $N$
independent first order equations. Hence, there are two more
equations on the WZNW side than on the Liouville side. Closer
inspection shows that the $N-2$ BPZ differential equations
correspond inserting the Sugawara tensor at the $N-2$ points
$w=y_i$ on the world-sheet of the WZNW model. At these special
points, the differential equations from the Sugawara singular
vector simplify considerably. To see this we need the operator
product expansions between the currents $J^\pm$ and $J^0$ and the
vertex operators in the $\mu$ basis \eqref{opejv}.
{}From the
current Ward identities and the very definition of the points
$y_i$ we conclude that
$$ J^-(y_i)\ = \ \sum_{\nu=1}^N \ \frac{\mu_\nu}{y_i - z_\nu}
\ = \ 0\ \ . $$ By similar reasoning, the current $J^0(y_i) $ is
easily seen to simplify as follows
$$  J^0(y_i) \ = \ - \sum_{\nu=1}^N
  \frac{\mu_\nu }{y_i - z_\nu}\ \frac{\pl}{\pl \mu_\nu} \ = -
\frac{\partial}{\partial y_i} \ \
$$
where the second equality follows from the explicit chance
of variables relating $\mu_\nu$ with $y_i$ and $u$. Once we
insert all these results into the Sugawara singular vector,
we obtain the following $N-2$ differential equations
\begin{equation} \label{WZNWeq0}
D^{\KZ}_i \Omega^H(z_\nu,\mu_\nu) \ = \ 0  \ \ \ \
 \mbox{ where } \ \  \Omega^H (z_\nu,\mu_\nu) \ = \
\blangle \ \prod_{\nu=1}^{N} \, V_{j_\nu}(\mu_\nu|z_\nu)
\ \brangle^H \ \ .
\end{equation}
Using the relation (\ref{mufromuy}), the differential operators 
$D^{H}_i$ for $H^+_3$ can be shown to acquire the following form
\begin{equation}
D^{\KZ}_i \ =\
b^2 \frac{\partial}{\partial y_i^2 } +
\sum_{\nu=1}^N \left(
 \frac{1}{y_i - z_\nu}\, \pl_\nu
  + \frac{\Delta^{H}_\nu}{(y_i - z_\nu)^2}  \right) ~,
\end{equation}
where $\Delta^{H}_\nu = - b^2 j_\nu (j_\nu + 1)$ are the
conformal dimensions of our WZNW primary fields, as before.
Note that $D_i^{\KZ}$ does not depend on any of the $y_j$ 
with $j \neq i$. In this sense, the transformation 
(\ref{mufromuy}) leads to a {\em separation of variables} 
\cite{Sklyanin}.  
Let us stress that the derivatives $\partial_\nu$ in eq.\
(\ref{WZNWeq0}) are to be taken with fixed $\mu_\nu$ whereas
the variables $y_i$ are kept fixed when evaluating
$\partial_\nu$ in the BPZ differential equations above.
Whenever it is relevant, we shall explicitly distinguish
between the two derivatives $\partial_\nu$,
$$ \partial^{\KZ}_\nu \ = \ \left(
   \frac{\partial}{\partial z_\nu}\right)_\mu \ \ \ , \ \ \
  \partial^{\BPZ}_\nu \ = \ \left(
   \frac{\partial}{\partial z_\nu}\right)_y\ \ .
$$
It is certainly possible to express e.g.\ all the derivatives
$\pl_\nu^{\KZ}$ in terms of $\pl^{\BPZ}_\nu, \partial_i$
and $\partial/\partial u$. We shall only need the special
combination
\begin{equation}
 \delta_i \ := \ \sum_\nu \, \frac{1}{y_i - z_\nu}\,
   \pl^{\KZ}_\nu
 \ = \  \sum_{\nu}\frac{1}{y_i - z_\nu}
 \left( \partial_\nu^\BPZ + \partial_i \right)
 - \sum_{j \neq i }\frac{1}{y_i - y_j}
 \left( \partial_i
      -  \partial_j \right) \ \ .
\end{equation}
With this auxiliary formula it is then straightforward to check
that
\begin{equation}
 \left( \,  \Theta_N^{-1} D^{\KZ}_i \Theta_N - D^{\BPZ}_i
 \right) \, \Omega^L (z_\nu , y _i )\ = \ 0 \ .
\end{equation}
The correspondence between differential equation was an important
ingredient in \cite{RT} for  proving the relation \eqref{corrg0}.
We have derived the latter within the path integral approach and
therefore the differential equations are guaranteed to be mapped
onto each other.

\subsection{Comments on the path integral derivation}

It is well known that the path integral definition of both the
WZNW model and of Liouville theory has some issues that are
related to the non-compactness of the background (see e.g.
\cite{Srv} for a review). If one splits all fields into their zero
modes and fluctuations, one can integrate out the zero modes. The
procedure results in expression for the correlators which are
either divergent or hard to give an exact definition. In the case
of Liouville theory, for instance, one obtains
$$
\langle \prod_{\nu=1}^N \,
     e^{2\alpha_\nu \varphi(z_\nu,\bz_\nu)} \rangle^L \ = \
     \int {\cal D} \tilde \varphi \, e^{-S_{\text{LD}}[\tilde \varphi]} \,
    \prod_{\nu=1}^N \, e^{2\alpha_\nu \tilde \varphi(z_\nu,\bz_\nu)}
   \frac{\Gamma(-s)}{2b} \, \left( \mu_B \int d^2w e^{2b \tilde
   \varphi} \right)^s
$$
where $sb =  Q_\varphi  - \sum_{\nu=1}^N \, \alpha_\nu$. In this
formula, the integration over the fluctuation field $\tilde
\varphi$ is weighted with the measure of the free linear dilaton
theory, but the integrand contains additional insertions of
screening charges. The latter are raised to some possibly
non-integer power $s$ and they are multiplied with coefficients
that diverge for positive integer powers. These features have a simple
explanation. In our non-compact target space, the correlation
functions are expected to possess poles in momentum space which
come from the integration over the infinite region of the target
space where the interaction is negligible. The path integral
computation we have just sketched detects (some of) these
divergencies and can be turned into a rigorous computation of the
associated residues.
\smallskip

If we are not willing to give the path integral any more
credit, then our results above only imply that
\begin{align} \label{Scomp}
 \blangle \ \prod_{\nu=1}^{N} \, V_{j_\nu}(\mu_\nu|z_\nu) \
   S^s_\phi\  \brangle^{\b \c}_{LD^\phi} & = \
  \delta^2 (\sum_{\nu=1}^N \mu_\nu)
|\Theta_N(u,y_i,z_\nu)|^2 \, \times \\[2mm]
 &  \hspace*{2cm}  \times \ \blangle \ \prod_{\nu=1}^N \,
V_{\alpha_\nu} (z_\nu) \ \prod_{i=1}^{N-2} \,
V_{-1/2b}(y_i)\  S^s_\varphi\  \brangle_{LD^\varphi}\ \ . \nonumber
\end{align}
Here, $S^s_\phi$ and $S^s_\varphi$ denote the screening charges
of the WZNW model and Liouville field theory, respectively.
They are given by the following expressions
\begin{equation} \label{screening}
S^s_\phi \ = \ - \int d^2w \ \b(w)\bb(\bar w) \, e^{2b\phi(w,\bar w) }
 \ \ \ \ , \ \ \ \
S^s_\varphi \ = \ \int d^2 w \ e^{2b\varphi(w,\bar w)} ~.
\end{equation}
The correlation functions on both sides of the equality are
to be computed in the free linear dilaton theory. On the left hand
side, we use the $\b\c$ system with central charge $c=2$ and a
linear dilaton with background charge $Q_\phi = b$. On the
right hand side, the correlator is to be computed for a
linear dilaton background with $Q_\varphi = b+ 1/b$.
\smallskip

Correlation functions in the WZNW model and in Liouville field
theory are well known to possess a second series of poles that
are not explained by insertions of the screening charges
(\ref{screening}). The residues of these poles can still be
computed from free field theory with the help of so-called
dual screening charges. For our two models these read
\begin{equation} \label{Dscreening}
S^d_\phi \ = \ - \int d^2w \ \b(w)^{\frac{1}{b^2}}
    \bb(\bar w)^\frac{1}{b^2} \, e^{\frac{2}{b}\phi(w,\bar w) }
 \ \ \ \ , \ \ \ \
S^d_\varphi \ = \ \int d^2 w \ e^{\frac{2}{b}\varphi(w,\bar w)}\ .
\end{equation}
Note that the exponents $1/b^2 = k-2$ can be integer for integer
level $k \geq 2$. We have indeed checked by explicit computation
that equation (\ref{Scomp}) remains valid if the screening charges
$S^s_\phi$ and $S^s_\varphi$ are replaced by the dual ones. Similar
calculations for the $H^+_3$ model involving both screening charges
$S^s_{\phi}$ and $S^d_{\phi}$ can be found, e.g., in \cite{GN} 
(see also references therein for earlier works on this subject). 
We shall perform free field computations for correlators 
on surfaces of genus $g$ at the end of section 4.

\section{Generalizing the correspondence to the torus}

Motivated by our rather simple path integral derivation, we would
now like to extend the $H^+_3$-Liouville relation
beyond tree level to higher genus correlators. We shall study the
general case in the next section and restrict ourselves to the
torus for now since many of the formulas can be made very
explicit at $g=1$.

\subsection{The $H_3^+$ model on the torus}

We start from the $H^+_3$ WZNW model with level $k$ and compute
$N$-point functions on a torus with moduli parameter $\tau$ using
the first order formulation in terms of $\phi$, $\beta$, $\gamma$.
As explained in the introduction, we would like to introduce a bit
more freedom by admitting some non-trivial boundary conditions for
$\phi$ and the $\b \c$ system. To be precise, we assume the fields
$\phi$, $\c$ and $\b$ to satisfy
\begin{align}
 \beta (w + m + n \tau ) & = \ e^{ 2 \pi i n \lambda} \beta (w) ~,
 \qquad \qquad
 \gamma (w + m + n \tau ) \ = \ e^{- 2 \pi i n \lambda} \gamma (w) ~,
 \label{bct1} \\[3mm]
 &\phi (w+m + n \tau, \bar w+m + n \bar \tau)
 \ = \ \phi (w) + \frac{2 \pi n {\rm Im } \lambda}{b}
  \label{bct2}
\end{align}
with $n,m \in {\mathbb Z}$ and $\lambda$ some complex parameter.
Once such twisted boundary conditions have been introduced for the
field $\b$, the conditions on $\c$ and $\phi$ follow if we require
the action to be single valued. Of course, we assume similar
twisted boundary conditions with twist parameter $\bar \lambda$ to
hold for the anti-holomorphic components of the $\b\c$ system.
\smallskip

There is one term in the action, namely the coupling of the field
$\phi$ to the world-sheet curvature, that requires a bit of
additional care. Since our field $\phi$ is multivalued, the term
$\sqrt{g} {\cal R}\, \phi$ cannot possibly give the right
prescription. Instead, we must decompose $\phi$ into a twisted
zero mode part $\phi_{\text{sol}}$ (the index `sol' stands for
{\em{solitonic}}) and a doubly periodic fluctuation
$\phi_q$ , i.e.\
\begin{align} \label{phi0}
\phi \ = \ \phi_q + \phi_{\text{sol}}  \ \ \ \ \mbox{ where }\ \ \
\phi_{\text{sol}} (w,\bar w) \ = \ \frac{2 \pi}{\tau_2 b} {\rm Im}
\lambda {\rm Im} w ~ \ ,
\end{align}
where, by construction, $\phi_q$ now satisfies $\phi_q (w+m + n \tau,
\bar w+m + n \bar \tau) \ = \ \phi_q (w)$.
The linear dilaton term in the action couples the world-sheet
curvature to the single valued fluctuation field $\phi_q$ rather
than to $\phi$ itself, \begin{equation}  S_{\text{LD}} [\phi] \ = \
\frac{Q_\phi}{8\pi} \int d^2w \,
 \sqrt{g} {\cal R}\, \phi_q\ \ . \label{LD} \end{equation}
All other terms in the action (\ref{action}) remain the same as
before. Similarly, the expression for vertex operators
(\ref{vertex}) does not require any modification.
\smallskip

To a large extend the path integral computation of the $N$-point
correlation function follows the same steps as before. The
quantities we would like to compute are given by
$$ \blangle \ \prod_{\nu=1}^{N} \, V_{j_\nu}(\mu_\nu|z_\nu) \
\brangle^H_{(\lambda,{\tau})} \ = \
 \int {\cal{D}^\lambda} \phi {\cal{D}^\lambda}
\c {\cal{D}^\lambda} \b \ e^{-S[\phi,\c,\b]} \prod_{\nu=1}^{N} \,
   V_{j_\nu}(\mu_\nu |z_\nu)
\ , $$ where the integration is to be performed over all field
configurations on a torus with modulus $\tau$ satisfying the
boundary conditions (\ref{bct1}) and (\ref{bct2}) stated above.
Note that our correlation function is not yet normalized by
dividing the partition function $Z^H$. We shall further comment on
this below. After integration over $\c$ and $\bc$ we
obtain the same condition (\ref{plb}) for the derivatives
of $\b$ and $\bb$ on the sphere. But this time it has different
consequences for $\b$ and $\bb$. In fact, we need a bit of
preparation before we are able to spell out the analogue
of the important equation (\ref{bspb}).
\smallskip

On the torus, the integration of equation (\ref{plb}) will lead to a
new function that can be constructed out of Jacobi's theta
function
\begin{align}
 \theta (z) &= \ - \sum_{n \in {\mathbb Z}}
 e^{i \pi (n+ \frac{1}{2})^2 \tau + 2 \pi i (n+ \frac{1}{2})
 (z + \frac{1}{2})} ~,
 &\theta (z + m + n \tau) &= - e^{- i \pi n ( 2 z + n \tau )}
\theta (z ) ~.
\end{align}
Out of $\theta$ we can build a new function $\sigma_\lambda$ with
a simple pole and the same periodicity properties that we required
for $\b$,
\begin{equation}
 \sigma_{\lambda} (z, w) \ = \  \frac{\theta (\lambda -(z-w) ) \theta ' (0)}
                              {\theta (z-w) \theta (\lambda ) } \ \ .
\end{equation}
Indeed, from the shift properties of Jacobi's theta function
$\theta$ it is easy to derive the following property of
$\sigma_\lambda$,
\begin{equation}
\sigma_{\lambda} (z + m + n \tau ,w) \ = \ e^{2 \pi i n \lambda}
 \sigma_{\lambda} (z,w) ~.
\end{equation}
Now let us return to the integration of the two equations
(\ref{plb}). The right hand side tells us that the twisted
meromorphic differential $\beta(w)$ possesses $N$ poles in the
positions $w=z_\nu$ with residues $ \mu_\nu$. The
solution to these conditions is unique, as long as the twist
parameter $\lambda$ does not vanish. It can be written down in
terms of the $\sigma_\lambda$ as
\begin{align}
 \beta(w) \ = \ \sum_{\nu=1}^N  \mu_\nu
\sigma_{ \lambda} (w , z_\nu)
  \ = \ u
 \frac{\prod_{i=1}^{N} \theta (w - y_i)}
      {\prod_{\nu=1}^N \theta (w - z_\nu) }
      \ = : \ u \, {\cal B}_1(y_i,z_\nu;w) ~.
\label{relation}
\end{align}
The second equality is a torus version of Sklyanin's separation of
variables and it requires a few comments. On a torus, meromorphic
one-differen\-tials possess the same number $N$ of poles and
zeroes. Moreover, the positions $w= z_\nu$ and $w= y_i, i=1,
\dots,N,$ of both zeroes and poles determine the differential up
to an overall factor $u$. For our correspondence between the WZNW
model and Liouville theory, it is again crucial to parametrize
$\rho \beta$ through $u$ and $y_i$ rather than $\lambda$ and $\mu_\nu$.
The relation between the two sets of parameters can be worked out
easily (see also \cite{EFR,FS})
\begin{equation}
 \mu_\nu \ = \ u \frac{\prod_{i=1}^N
  \theta (z_\nu - y_i)} {\theta ' (0) \prod_{\mu \neq \nu, \mu  = 1}^N
  \theta (z_\nu - z_\mu)} \ \ \ , \ \ \ \
  \lambda  \ = \ \sum_{i=1}^N y_i - \sum_{\nu=1}^N z_\nu ~ .
 \label{rel1}
\end{equation}
What we have shown so far is that the integration over the fields
$\gamma$ and $\bar \gamma$ in the WZNW model leads to the
following $\vardelta$ function
\begin{equation}\label{delta1}
\vardelta^2 (\bar \partial \beta (w) -
    2\pi \sum_{\nu=1}^N \mu_\nu \delta^2(w-z_\nu) ) \ = \
    |{\rm det } \partial|^{-2}_\lambda
 \vardelta^2 ( \b(w) - u\,  {\cal B}_1 (y_i,z_\nu;w))\ \ .
\end{equation}
This replaces the related formula (\ref{delta0}) in the genus zero
analysis. The factor $ |\det\partial|^{-2}_\lambda$ is the
Jacobian that arises when we change from $\delta^2(\partial \beta
\cdots)$ to $\delta^2 ( \b(w) \cdots)$. We have placed a subscript
$\lambda$ in the Jacobian  to remind us that $\partial$ is
considered as an operator on twisted one-differentials. Let us now
perform the integration over $\beta$ and $\bar \beta$ to obtain
\begin{eqnarray*}
 \blangle \ \prod_{\nu=1}^{N} \, V_{j_\nu}(\mu_\nu|z_\nu) \
 \brangle^H_{(\lambda,\tau)}
& = & \frac{1}{|\det \partial |^{2}_\lambda}
 \int {\cal{D}^\lambda} \phi e^{-\frac{1}{2\pi} \int d^2w \ \left(
\bpl \phi \pl \phi + \frac{Q_\phi}{4} \sqrt{g} {\cal R} \phi _q + b^2
|{\mathcal{B}_1}|^2\, e^{2 b \phi}\right) }\,
\times \\[2mm] & & \hspace*{0cm} \times
\prod_{\nu=1}^{N} |\rho(z_\nu)|^{2\Delta^J_\nu} \,
|u|^{-2(j_\nu+1)} |\mu_\nu|^{ 2(j_\nu + 1)} e^{2b (j_\nu +1 )
\phi(z_\nu)}\ .
\end{eqnarray*}
As in our genus zero analysis we have shifted the zero mode of the
field $\phi$ to remove the $|u|^2$ from the interaction term.
Because the Euler characteristics of the torus vanishes, the path
integral is multiplied with a factor $|u|^{0} = 1$.
\smallskip

We have now reached the point at which we change variables
for the remaining integration over $\phi$ such that the highly
non-trivial factor $|{\cal B}_1|^2$ also gets removed from the
interaction. In complete analogy to the genus zero case we
introduce
\begin{equation}
\varphi(w,\bar w) \ := \ \phi(w,\bar w) + \frac{1}{2 b} \left(
\sum_i \ln |\theta ( w - y_i)|^2
 - \sum_\nu \ln |\theta (w - z_\nu )|^2 - \ln |\rho(w)|^2 \right) \ \ .
\end{equation}
By construction, the new field $\varphi$ is periodic under shift
by $n+m\tau$, even though the original field $\phi$ and the
$\theta$ functions are not. It will be advantageous to replace
$|\theta|^2$ functions by some function $F$,
\begin{equation}
 F(z,w) \ := \ e^{- \frac{2 \pi}{\tau_2} ( {\rm Im} (z-w))^2  }
  \left| \frac{\theta ( z - w ) }{\theta ' (0) }\right|^2 ~,
\end{equation}
which is easily seen to be invariant under shifts by integers and
integer multiples of the modulus $\tau$,
$$
 F(z + m + n \tau,w) \ = \ F(z,w) \ \ \ .
$$
The new function $F$ allows us to rewrite the relation between
$\varphi$ and $\phi$ in terms of the single valued fluctuation
field $\phi_q = \phi-\phi_{\text{sol}}$ that we introduced in eq.\
(\ref{phi0}),
\begin{align}
\varphi(w,\bar w) := \phi_q (w,\bar w) + \frac{1}{2 b} \left(
\sum_i \ln F ( w ,y_i)
 - \sum_\nu \ln F (w,z_\nu ) - \ln |\rho(w)|^2 + S_1 \right)
\end{align}
with $S_1 = \frac{2 \pi }{\tau_2} ( \sum_i y_i^2  - \sum_\nu
z_\nu^2)$. Let us also observe that the decomposition of the field
$\phi$ into $\phi_{\text{sol}}$ and $\phi_q$ is such that their
contributions to the kinetic term decouple,
\begin{align} \label{splitkin}
 - \frac{1}{2\pi} \int d^2 w \partial \phi \bar \partial \phi
 & = \ - \frac{1}{2\pi} \int d^2 w \partial \phi_{\text{sol}} \bar
  \partial \phi_{\text{sol}}
 - \frac{1}{2\pi} \int d^2 w \partial \phi_q \bar
\partial \phi_q\ \\[2mm]
\mbox{with}\ \ \ \ \  &  \frac{1}{2\pi} \int d^2 w \partial
\phi_{\text{sol}} \bar  \partial \phi_{\text{sol}}\ = \
\frac{\pi ({\text{Im}} \lambda)^2}{b^2 \tau_2} \ \ .
\end{align}
Now we can proceed exactly as before, with the function $F$
replacing $|z-w|^2$. In particular, the properties of $F$ can
be used to evaluate $\pl \bpl \varphi$ as
\begin{equation} \label{plbplphi}
\pl \bpl \varphi(w,\bar w) := \pl \bpl \phi_q (w,\bar w) +
\frac{\pi}{b} \left(\sum_i \delta^2(w-y_i) - \sum_\nu
\delta^2(w-z_\nu) \right) - \frac{1}{2b} \pl\bpl \ln|\rho(w)|^2,
\end{equation}
a result that agrees exactly with the corresponding formula at
genus zero. The outcome of a short and straightforward computation
is
\begin{align*}
\blangle \, \prod_{\nu=1}^{N} \, V_{j_\nu}(\mu_\nu|z_\nu) \,
\brangle^H_{(\lambda,\tau)}
 &= e^{- \frac{\pi ({\rm Im} \lambda)^2}{b^2 \tau_2}}
 \frac{|\Theta^{g=1}_N(y_i,z_\nu,\tau)|^2 }{|\det \partial |^2_\lambda}
  \int {\cal{D}}
\varphi \, e^{-\frac{1}{2\pi}
\int d^2w \, \left( \bpl \varphi \pl \varphi
 + \frac{Q_{\varphi}}{4 }\sqrt{g} \mathcal{R} \varphi + b^2 |\rho|^2 e^{2 b
\varphi}\right) }\  \\ & \times \
\prod_{\nu=1}^{N} \, |\rho (z_\nu)|^{2 \Delta^L_{\nu}}
e^{2\left( b (j_\nu +1) +
\frac{1}{2b}\right) \varphi(z_\nu)} \, \prod_{i=1}^{N} \,
 |\rho (y_i)|^{2 \Delta^L_{-1/2b}}
e^{-\frac{1}{b} \varphi(y_i)}\ \
\end{align*}
where the pre-factor $\Theta_N$ is given by
\begin{align}
 |\Theta^{g=1}_N (y_i,z_\nu,\tau)|^2
  \ = \
     \prod_{\mu < \nu}^N F ( z_\mu , z_\nu )^{\frac{1}{2 b^2}}
      \prod_{i < j}^{N} F (y_i , y_j )^{\frac{1}{2 b^2}}
      \prod_{\mu,i=1}^N
       F (z_\mu , y_i)^{- \frac{1}{2 b^2}} ~.
\end{align}
So far, the correlators on both sides of the equations are not
normalized. But using the following formulas (see \cite{Gawedzki}
and our discussion before eq.\ (\ref{partitionfct}) later on)
for the genus one partition functions $Z^H$ and $Z^L$ of the WZNW
model and Liouville theory, respectively,
\begin{align}
Z^H \ = \ \frac{e^{ - \frac{\pi ( {\rm Im} \lambda ) ^2}{b^2
\tau_2} }} {\sqrt{\tau_2} | \theta (\lambda)|^2} \ \ \ \ , \ \ \ \
\ Z^L \ = \ \frac{1}{\sqrt{\tau_2} | \eta (\tau ) | ^2}\ \ ,
\end{align}
along with the formula $\det \partial_\lambda =
\theta(\lambda)/\eta(q)$ for the determinant of $\partial$ on
$\lambda$-twisted differentials, we can recast the relation
between correlation functions in both theories in a particularly
simple form
\begin{align}
\frac{1}{Z^H} \blangle \ \prod_{\nu=1}^{N} \,
V_{j_\nu}(\mu_\nu|z_\nu) \
   \brangle^H_{(\lambda,\tau)} \ = \
|\Theta^{g=1}_N(y_i,z_\nu,\tau)|^2 \, \frac{1}{Z^L} \blangle \,
\prod_{\nu=1}^N \, V_{\alpha_\nu} (z_\nu) \ \prod_{i=1}^{N} \,
V_{-1/2b}(y_i)\, \brangle^{L}\ \ . \label{corrg1}
\end{align}
Our genus one relation is similar to the tree level result. Once
more we managed to compute all correlators of WZNW primaries in
terms of Liouville theory with the same relation between the level
$k = b^{-2} + 2$ and the background charge $Q_\varphi = b + 1/b$
and the same bulk cosmological constant $\mu_B = 4 b^2$. For each
primary field $V_j$ in the WZNW model there appears one Liouville
vertex operator $V_\alpha$ where $\alpha = b(j+1) + 1/2b$, as on
the sphere. Additionally we have to insert degenerate Liouville
fields, but now we need two more than on the sphere, i.e.\ there
are $N$ degenerate fields inserted. Their positions are determined
through the light cone momenta $\mu_\nu$ of the WZNW vertex
operators and the twist parameter $\lambda$.
\smallskip

The attentive reader might be a bit surprised not to see any
factor implementing the conservation of $\mu$ momentum, as on the
sphere. Its absence is directly related to the fact that the
correspondence has been worked out for non-zero twist parameter
$\lambda$. One would certainly expect to recover $\mu$ momentum
conservation in the limit $\lambda \rightarrow 0$. In order to see
how this works, let us go back to relation (\ref{relation}) and
insert the expansion
\begin{equation}
\sigma_\lambda(z,w) \ = \ \frac{1}{\lambda} +
  \frac{\theta'(w-z)}{\theta(w-z)} + \dots\ .
\end{equation}
This leads to the formula
$$ \beta(w) \ = \ \sum_\nu \, \frac{\mu_\nu}{\lambda} +
  \sum_\nu \, \mu_\nu \, \partial_w \ln \theta(w-z_\nu) + \dots \
  .
$$
For $\beta$ to stay finite in the limit $\lambda \rightarrow 0$,
we need to assume that the total $\mu$-momentum $\sum_\nu \mu_\nu$
tends to zero when we send $\lambda \rightarrow 0$. In fact, if
$\sum_\nu \mu_\nu$ vanishes fast enough, we obtain well defined
expressions at $\lambda = 0$.

\subsection{Relation between differential equations on torus}

Having established a simple relation between correlation functions
of the WZNW model and Liouville field theory on the torus it seems
worthwhile to look once more at the differential equations that
determine correlators in both models and to check that our
relation (\ref{corrg1}) correctly intertwines between them.

Let us start on the side of Liouville field theory. The vertex
operator $V_{-1/2b}$ belongs to a degenerate representation with
a null vector $(b^2 (L_{-1})^2 + L_{-2} )|-1/2b\rangle$ on the
second level. Since we have $N$ such degenerate fields in our
Liouville correlation function, we obtain $N$ second order
differential equations,
\begin{align}
 D^{\BPZ}_i (\Omega^L ) (z_\nu , y_j , \tau ) &= \ 0 ~,
 &\Omega^L(z_\nu,y_i,\tau) &= \  \blangle
 \prod_{\nu=1}^N V_{\alpha_\nu} (z_\nu ) \prod_{j =1}^N
 V_{-1/2b} (y_j) \brangle^L  ~ .
 \label{BPZ}
\end{align}
The differential operators $D^{\BPZ}_i$ for surfaces of higher
genus were worked out by Eguchi and Ooguri \cite{EO},
\begin{align}
 D^{\BPZ}_i \ = \ b^2 \frac{\partial^2}{\partial y_i^2}
  + \Delta^L_{- \frac{1}{2b}} 2 \eta_1
  + \sum_{j \neq i} \left( \xi (y_i , y_j )
 \frac{\partial}{\partial y_j}
 - \Delta^L_{ - \frac{1}{2b}} \partial_i \xi (y_i , y_j) \right) +
 \nonumber \\[2mm]
   + \sum_{\nu} \left( \xi (y_i , z_\nu )
 \frac{\partial}{\partial z_\nu}
 - \Delta^L_{ \alpha_\nu} \partial_i \xi (y_i , z_\nu) \right)
 + 2 \pi i \frac{\partial}{\partial \tau}\  ~,
 \label{BPZop}
\end{align}
where the special function $\xi$ and the constant $\eta_1$ are
constructed from Jacobi's $\theta$ function through
\begin{equation} \label{eta1}
  \xi ( z , w )  = \ \frac{\theta ' (z - w ) }{\theta (z - w)} ~,
  \qquad \eta_1 =\  - \frac{1}{6}
 \frac{\theta ^{'''}(0)}{\theta ^{'} (0)}\ ~.
\end{equation}
Let us now address the differential equations obeyed by the
correlation functions of the WZNW model on the torus, which were
first worked out by Bernard and are known as
Knizhnik-Zamolodchikov-Bernard (KZB) equations. These equations are
obtained by inserting the Sugawara singular vector $T(w) - b^2 :
J^a J_a : (w) = 0$ into the WZNW $N$-point correlation function
$\Omega^H$ (see eq.\ (\ref{omdef}) below for the definition of
$\Omega^H$). Ward identities for currents and the Virasoro field
then give \cite{Bernald,Bernald2}
\begin{align}
  \left[ b^2 S (w)
+ \sum_{\nu =1}^N \left( \xi (y_a,z_\nu)
 \frac{\partial}{\partial z_\nu} - \Delta^H_{j_\nu} \partial_i \xi (y_i,z_\nu)
 \right) + 2 \pi i \frac{\partial}{\partial \tau} \right]
  \Pi \Omega^H  = 0 ~.
\end{align}
Here, we introduced $\Pi = |\theta(\lambda)|^2$ and
\begin{align}
 S(w) &=\
\left( \frac{\partial}{\partial \lambda} - \tilde
 J ^0(w) \right)^2 - \frac{1}{2}
  \left(  \tilde J^-(w) \tilde J^+(w)+
\tilde J^+(w) \tilde J^-(w) \right) ~ ,
\end{align}
\begin{equation}
  \tilde J^{\mp} (w ) = \sum_{\nu=1}^N\,  \sigma_{\pm \lambda}
   (w , z_\nu)\, D^{\mp}_\nu~\ \ \ , \qquad
\tilde J^0 (w ) = \sum_{\nu=1}^N \, \xi (w ,z_\nu )\,  D^0_\nu\   ~,
\end{equation}
and the differential operators $D_\nu^\pm$ and $D^0_\nu$ are the
same as in eq.\ (\ref{mubasis}). Let us briefly recall the reason
why the KZB equations contain a derivative with respect to the
twist parameter $\lambda$. These terms arise from the Ward
identities of currents. In fact, it has already been observed by
Eguchi and Ooguri in \cite{EO} that the insertion of the zero
modes of currents into correlators cannot be converted into
differential operators acting on the usual untwisted correlation
functions. It was Bernard's idea to fix this problem by
introducing a dependence of conformal blocks on additional
parameters. On the torus, he suggested to insert a group element
$g$ into the trace. This has the effect of twisting the boundary
conditions for currents under shifts by multiples of $\tau$. Our
boundary conditions correspond to the special choice 
$g = \exp (- 2 \pi i \lambda J^0_0)$. 
Actually, it had been observed by Bernard already
that a single twist parameter $\lambda$ suffices on the torus.
\smallskip

As in subsection 2.2, our strategy now is to evaluate the KZB
equations at the $N$ special points $y_i$ and then to compare the
result with the $N$ differential equations for the Liouville
correlator. From the relation (\ref{rel1}) between $\mu_i,\lambda$
and $y_i,u$ one may derive
\begin{equation}
 \frac{\partial}{\partial y_i} = \ \frac{\partial}{\partial \lambda}
  + \sum_{\nu=1}^N\,
  \xi(y_i,z_\nu) \,
  \mu_\nu \frac{\partial}{\partial \mu_\nu}
   ~,  \qquad u \frac{\partial}{\partial u}
  = \sum_{\nu=1}^N \mu_\nu \frac{\partial}{\partial \mu_\nu} ~.
 \label{op}
\end{equation}
These relations between derivatives can be inserted into our
formulas for the differential operators $\tilde J^{\pm}(w)$ and
$\tilde J^{0}(w)$, evaluated at the points $w=y_i$, to obtain
$$
 \tilde J^- ( y_i ) \ = \ \sum_{\nu=1}^N \sigma_{+ \lambda}
   (y_i - z_\nu) \mu_\nu = 0 ~\ \ ,
\ \ \  \tilde J^0 ( y_i ) \ = \ - \sum_{\nu=1}^N \xi (y_i ,z_\nu )
 \mu_\nu \frac{\partial}{\partial \mu_\nu}
 = - \frac{\partial}{\partial y_i}
 + \frac{\partial}{\partial \lambda}\ \ .
$$
When we plug these expressions into $S(y_i)$ we find
\begin{align}
 S(y_i) \ = \ \frac{\partial ^2}{\partial y_i ^2} ~,
\end{align}
just as on the sphere. In conclusion we have shown that the KZB
equations lead to the following $N$ differential equations for the
WZNW $N$-point functions $\Omega^H$
\begin{align} \label{omdef}
  &D^{\KZB}_i  \Pi \Omega^H \ = \ 0 \quad ~, \quad \  \Omega^H (z_\nu ,
  \mu_\nu,\lambda)\ =\
  \blangle \prod_{i = 1}^N V_{j_i} (\mu_i | z_i ) \brangle ^H_{(\lambda,\tau)}
  ~,\\[2mm]
  &D^{\KZB}_i \ = \  b^2 \frac{\partial ^2}{\partial y_i ^2}
+ \sum_{\nu =1}^N \left( \xi (y_i,z_\nu)
 \frac{\partial}{\partial z_\nu} - \Delta^H_{j_\nu} \partial_i \xi (y_i,z_\nu)
 \right) + 2 \pi i \frac{\partial}{\partial \tau} ~\ \ .
 \label{KZBop}
\end{align}
Let us recall that the derivatives $\partial/\partial z_\nu =
\partial_\nu$ in $D^{\KZB}_i$ are still taken while keeping $\lambda$
and $\mu_i$ fixed, in spite of the explicit appearance of
derivatives with respect $y_i$.

In order to verify consistency of the two sets of equations with
the proposed relation between Liouville and WZNW correlation
functions, we rewrite the latter in the form
\begin{align}
  & \Pi \Omega^H \ = \ |\eta (\tau) |^2 | \Theta_N ' |^2
 \Omega^L ~, \label{corres} \\[2mm]
  &\Theta_N ' =\
 \theta '(0) ^{\frac{N}{2 b^2}}
      \prod_{\nu < \mu}^N  \theta ( z_\nu - z_\mu ) ^{\frac{1}{2 b^2}}
      \prod_{i < j}^{N}  \theta (y_i - y_j ) ^{\frac{1}{2 b^2}}
      \prod_{\nu,i=1}^N  \theta (z_\nu - y_i) ^{- \frac{1}{2 b^2}} ~.
  \label{twistg1}
\end{align}
In comparison to the earlier version, we have absorbed a factor
into a re-definition of $|\Theta_N|^2$ and then expressed the new
$|\Theta_N'|^2$ in terms of $|\theta|^2$ rather than $F$. Our result
on the relation between correlation functions therefore implies
\begin{align}
(\eta \Theta '_N   )^{-1} D_i^{\KZB} (\eta  \Theta ' _N )
 \ = \ D_i^{\BPZ}\ \    ~.
 \label{diffg1}
\end{align}
This can be checked indeed by a lengthy but straightforward
computation. In the process it is important to replace all the
derivatives $\partial_\nu = \partial_\nu^H$ in the differential
operators $D_i^H$ through derivatives $\partial_\nu =
\partial_\nu^L$ where the latter are taken while keeping $y_i$
and $u$ fixed. More precisely, we replace $\delta_i = \sum_\nu \xi
( y_i , z_\nu )\partial^H_\nu$ in $D^{\KZB}_i$ by
\begin{align} \label{deltai}
 \delta_i \ \equiv\  \sum_{\nu=1}^N \xi (y_i , z_\nu)
 \left(\partial_\nu^L
 +  \partial_i \right)
 -  \sum_{j \neq i} \xi (y_i , y_j)
 \left(\partial_i - \partial_j \right) ~\ \ ,
\end{align}
where $\partial_i$ denote differentiation with respect to $y_i$,
as before. It is easy to verify that $\delta_i$ satisfy $\delta_i
\mu_\nu(y_i,z_\nu,u) = 0$ and $\delta_i \lambda(y_i,z_\nu,u) = 0$.
Once this issue is taken care of, we can show that the
differential operators indeed satisfy \eqref{diffg1}. The details
of the computation are presented in appendix \ref{apg1}.

\section{Generalization to arbitrary genus}

Equipped with the experience from genus $g=0$ and $g=1$, we now
address the WZNW model on an arbitrary closed surface of genus
$g$. Most of the analysis follows the ideas of previous sections,
but the details require considerably more background concerning
differentials on higher genus surfaces. We provide the most
relevant details in Appendix \ref{riemann}. Since this section
contains our main new result, we shall derive it first through our
path integral arguments and then verify with the help of free
field computations where both sides of the proposed relation
possess the same residues. Comments on the relation between the
KZB and BPZ differential equations are deferred to the concluding
section.

\subsection{Path integral derivation}

Suppose we are given some compact Riemann surface $\Sigma$ with
moduli parametrized by the period matrix $\tau = (\tau_{ij})$.
Since our fields are going to be multivalued as in the torus case,
it is appropriate to pass to the universal cover $\tilde \Sigma$
of the surface right away. Using the famous Abel map, we embed
$\tilde \Sigma$ into $\mathbb{C}^g$. From now on, we shall think
of our fields as being defined on the image of the Abel map and
hence consider them as functions of $g$ complex coordinates $w_k,
k=1,\dots,g$. In close analogy to the genus $g=1$ case, we allow
for nontrivial twists along the $\beta$-cycles, i.e.\
\begin{align}
 & \beta ( w_k + \tau_{kl} n^l + m_k | \tau )
 \  =\  e^{2 \pi i n^l \lambda_l} \beta (w_k | \tau ) ~,
  \nonumber \\[3mm]
 & \gamma ( w_k + \tau_{kl} n^l + m_k | \tau )
  \ = \ e^{ - 2 \pi i n^l \lambda_l} \gamma (w_k | \tau ) ~,
\label{bcg} \\[2mm]
 & \phi ( w_k + \tau_{kl} n^l + m_k | \tau )
  \ = \ \phi (w_k | \tau ) + \frac{2 \pi n^l {\rm Im} \lambda_l}{b} ~.
\nonumber
\end{align}
The complex parameter $\lambda_k$ represents the twist along the
$\beta$-cycle $\beta_k$. Thereby, we have introduced $g$ complex
parameters.
\smallskip

As in our discussion of the theory on the torus, spelling out the
coupling of $\phi$ to the world-sheet curvature requires to split
$\phi$ into a twisted zero mode $\phi_{\text{sol}}$ and a single
valued fluctuation $\phi_q$. The twisted zero mode $\phi_{\text{sol}}$
is now given by
\begin{align}
 \phi_{\text{sol}} \ =  \ \frac{2 \pi}{b}\,  {\rm Im} \lambda^k\
 ({\rm Im} \tau)^{-1}_{kl} \ {\rm Im} w^l
  \ = \ \frac{2 \pi}{b}\,  {\rm Im} \lambda^k\
 ({\rm Im} \tau)^{-1}_{kl} \ {\rm Im} \int^w_{w_0} \omega^l ~
\end{align}
where $\omega^l, l=1.\dots,g,$  is a basis of holomorphic one-forms
and indices $l,k$ are raised and lowered with the trivial metric. The
linear dilaton term couples the world-sheet curvature ${\cal
R}$ to the doubly periodic fluctuation field $\phi_q  = \phi -
\phi_{\text{sol}}$ in the same way as on the torus, see eq.\
(\ref{LD}).
\smallskip

So far, setting up the path integral for WZNW correlators on a
surface of genus $g$ was a straightforward extension of the torus
case. But there is one important modification. As is well known,
the $\lambda$-twisted differentials $\rho \b$ and $\rho \bb$
possess $g-1$ zero modes. These give rise to $g-1$ additional
moduli if we decide to fix the value of the $\rho \b$ and $\rho
\bb$ zero modes and to extend our path integral only over the
remaining fluctuations. Our aim therefore is to compute the
following $N$-point correlation functions
$$ \blangle \ \prod_{\nu=1}^{N} \, V_{j_\nu}(\mu_\nu|z_\nu) \
\brangle^H_{(\lambda,\varpi,\tau)} \ = \ \int {\cal{D}^\lambda}
\phi {\cal{D}^\lambda} \c {\tilde{\cal{D}}^\lambda}
\b \ e^{-S[\phi,\c,\b]}
\prod_{\nu=1}^{N} \,
   V_{j_\nu}(\mu_\nu |z_\nu) $$
over a Riemann surface with genus $g$. The symbol $\tilde
{\cal{D}}^\lambda \beta$ reminds us not to integrate over $\b$ zero modes.
We parametrize the latter by $g-1$ coordinates $\varpi =
(\varpi_\sigma, \sigma = 1, \dots, g-1)$ and place an explicit
subscript $\varpi$ on the correlator. The physical correlation
functions may be recovered in principle through a finite
dimensional integral over $\varpi_\sigma$.
\smallskip

Integration over $\gamma$ leads to exactly the same expression
(\ref{plb}) for the derivative of $\b$ as on the sphere and torus.
But the corresponding $\b$ takes a different form. From the
knowledge of its derivative and the boundary conditions
(\ref{bcg}) we may conclude that $\beta$ must have the form
\begin{equation} \label{bspbg}
\beta (w) \ = \   \sum_{\nu=1}^N \mu_\nu \sigma_{\lambda} (w ,
z_\nu) + \sum_{\sigma=1}^{g-1} \varpi_\sigma \omega^\lambda_\sigma (w)
\end{equation}
where $\omega^\lambda_\sigma$ denote a basis of $\lambda$ twisted
holomorphic differentials and the function $\sigma_{\lambda} (w ,
z)$ is the following differential,
\begin{equation} \sigma_{\lambda} (w , z) \ =\
\frac{(h_{\delta}(w))^2}{\theta_{\delta} (\int^w_z \omega)}
\frac{\theta_\delta (\lambda - \int^w_z \omega)}
     {\theta_\delta (\lambda )} \ ~.
\end{equation}
On the right hand side, we can use any odd spin structure $\delta$.
The $\theta$ function $\theta_ \delta$ and the 1/2-differential
$h_\delta$ are defined in appendix \ref{riemann}. Using properties of
these objects it is possible to show that $\sigma_\lambda(z,w)$ has a
simple pole at $z=w$ with residue {\it Res}$_{z=w}
\sigma_\lambda(z,w) = 1$. Moreover the differential
$\sigma_\lambda$ satisfies the same periodic boundary condition as
$\beta (w)$. Before we go on, let us point out that we had to fix
the $\beta$ zero mode in order to be able to reconstruct $\beta$
from its derivative. Explicit formulas for the $g-1$ twisted
holomorphic differentials can be found e.g.\ in \cite{Bernald2}.
\smallskip

The rest of our analysis proceeds essentially as before. A
meromorphic one-differential with $N$ poles is known to possess
$N+2(g-1)$ zeroes. Hence, the analogue of the separation of
variables formula (\ref{bspb}) for a surface of genus $g \geq 1$
is given by
\begin{equation} \label{svg1}
 \b(w)
 \ =\
 \ u \,
\frac{\prod_{i=1}^{N + 2 (g-1)} E (w , y_i) \sigma (w)^2}
 {\prod_{\nu=1}^N E (w , z_\nu)}  \
 = : \ u \, {\cal B}_g(y_i,z_\nu;w) ~
\end{equation}
with $y_i$ parametrizing the zeroes of $\b$. It seems that 
this formula has not appeared in the literature before. In order 
to reproduce the correct boundary conditions \eqref{bcg},
$y_i$ have to satisfy the condition
\begin{align}
 \lambda_l = \sum_{k=1}^{N + 2 (g-1)} \int_w^{y_k} \omega_l
 - \sum_{\nu =1}^{N}   \int_w^{z_\nu} \omega_l
 - 2 \int_{(g-1) w}^{\Delta} \omega_l ~,
 \label{relg}
\end{align}
which corresponds to the condition \eqref{rel1} for the genus one
case. Notice that this condition does not depend on $w$. The
function $\sigma(w)$ in eq.\ (\ref{svg1}) is a $g/2$-differential
without zeros nor poles. Its definition is reviewed is eq.\ (\ref{hz})
of appendix \ref{riemann} along with the construction of the prime
form $E$ (see eq.\ (\ref{prime})) and the Riemann class $\Delta$.
The function $\sigma(w)$ is needed in order for the right hand
side of \eqref{svg1} to become a one-differential with the correct
zeros and poles.

There is another point we would like to stress. Note that the
equations (\ref{relg}) impose $g$ constraints on the position of
the zeroes $y_i$. Thereby, they define a $N+g-2$-dimensional
hyper-surface in the configuration space of $y_i$. These
hyper-surfaces sweep out the entire configuration space as we vary
the $g$ twist parameters. Changing the zero mode parameters
$\varpi_\sigma$ while keeping $\lambda$ fixed also moves the
position of zeroes $y_i$, but this motion takes place within the
the hyper-surface defined by $\lambda$ since eq.\ (\ref{relg}) is
independent of $\varpi$. Hence, the infinitesimal changes of the
twist parameters $\lambda$ and the zero mode parameters $\varpi$
span a $2g-1$ dimensional subspace of vectors tangent to the
configuration space of $y_i$. Together with the shifts of the $N$
light-cone momenta $\mu_\nu$, we thereby generate independent
moves of all the zeroes $y_i$ and of $u$. Let us also note that
the light-cone momenta may be reconstructed from $y_i$ according
to
\begin{equation} \label{muE}
 \mu_\nu \ =\  u \,
     \frac{\prod_{j=1}^{N + 2 (g-1)} E ( z_\nu , y_j) \sigma (z_\nu)^2}
          { \prod_{\mu \neq \nu}^N E (z_\nu , z_\mu)}\ .
\end{equation}
After these comments we can continue with our computation
of WZNW correlators. Once the trivial integration over non-zero
modes of $\b$ and $\bb$ has been performed, we re-define the
bosonic field,
\begin{align}
& \varphi(w,\bar w) \  := \nonumber\\ & \ \ \ \phi(w,\bar w) \,
+\, \frac{1}{2 b} \left( \sum_{i=1}^{N + 2 (g-1)} \ln |E ( w
,y_i)|^2
 - \sum_{\nu=1}^N \ln |E (w,z_\nu )|^2 + 2 \ln |\sigma (w)|^2
 - \ln | \rho(w)|^2 \right)\, . \nonumber
\end{align}
As on the torus, we may replace the multi-valued prime form $E$
and $\sigma$ by doubly periodic functions and thereby rewrite
$\varphi$ in terms of the single valued fluctuation field $\phi_q
= \phi -\phi_{\text{sol}} $,
\begin{align}
&\varphi(w,\bar w) \  = \nonumber \\
&  \ \ \  \phi_q (w,\bar w) \, + \, \frac{1}{2 b} \left(
\sum_{i=1}^{N + 2 (g-1)} \ln F ( w ,y_i)
 - \sum_{\nu=1}^N \ln F (w,z_\nu ) + 2 \ln H (w)
 - \ln | \rho(w)|^2 + S_g \right)\, . \nonumber
\end{align}
The functions $F(z),H(z)$ are defined in appendix \ref{riemann}
and the constant $S_g$ is a shorthand for the following expression
\begin{align}
 S_g \ = \ 2 \pi \sum_i\,  {\rm Im} \int^{y_i}_{w_0} \omega^l
 \ ({\rm Im} \tau)^{-1}_{lk} \ {\rm Im} \int^{y_i}_{w_0} \omega^k
  - 2 \pi \sum_\nu\,  {\rm Im} \int^{z_\nu}_{w_0} \omega^l
 \ ({\rm Im} \tau)^{-1}_{lk}\  {\rm Im} \int^{y_\nu}_{w_0} \omega^k
 -
 \nonumber \\[2mm]
  - 4 \pi {\rm Im} \int^{\Delta}_{(g-1) w_0} \omega^l
 \ ({\rm Im} \tau)^{-1}_{lk} \ {\rm Im} \int^{\Delta}_{(g-1)w_0}
 \omega^k
 ~, \nonumber
\end{align}
which is independent of $w_0$. If we act with $\pl \bpl$ on
$\varphi$ we obtain the same expression (\ref{plbplphi}) as on the
torus. This uses that the contributions from the non-holomorphic
part cancel each other. Furthermore, $\pl \bpl \ln \sigma (z) =0$
because $\sigma$ has neither zeros nor poles.
\smallskip

After inserting the shift of variables from $\phi$ to $\varphi$,
we can simplify the resulting expressions pretty much in the same
way as for the torus case. Once more, the kinetic term for the
field $\phi$ splits into the sum (\ref{splitkin}) of a constant
term and a kinetic term for the fluctuation field $\phi_q$. The
former is given by
\begin{align}
 S_{\rm sol} =  \frac{1}{2\pi} \int d^2w \,
  \bpl \phi_{\rm sol} \pl \phi_{\rm sol}
  = \frac{\pi}{b^2} {\rm Im} \lambda^l
\,  ({\rm Im} \tau)^{-1}_{lk}\,  {\rm Im} \lambda^k\  ~.
\end{align}
A second auxiliary result concerns the contribution from the
linear dilaton term. After the change of variables it provides us
with a linear dilaton term for $\varphi$ and the following
additional terms,
\begin{eqnarray*}
\frac{Q_{\phi}}{8 \pi} \int\! d^2 w \frac{\sqrt{g} \mathcal{R}}
  {2 b} \left( \sum_{i=1}^{N + 2 (g-1)}\!\!\!\ln F ( w ,y_i)
 - \sum_{\nu=1}^N \ln F (w,z_\nu ) + 2 \ln H(w) - \ln |\rho(w)|^2 +
 S_g\right) = \nonumber\\[2mm]
  = \sum_{i=1}^{N+2(g-1)} \ln ( |\rho (y_i)|^{-1} H (y_i) )
  - \sum_{\nu = 1}^N \ln ( |\rho (z_\nu)|^{-1} H(z_\nu) )
  + (1-g) S_g
 + \frac{3}{2} U_g ~,
\end{eqnarray*}
where (see \cite{Verlinde})
\begin{align}
 U_g \ = \ \frac{1}{192 \pi ^2 } \int d^2 w d^2 y \sqrt{g (w)} \mathcal{R}(w)
  \sqrt{g(y)} \mathcal{R}(y)
  \ln F(w,y)\   ~.
\end{align}
All these expressions can be verified using formulas from Appendix
\ref{riemann}. Collecting all the above facts, our  correlation
function is given by
$$
\blangle \ \prod_{\nu=1}^{N} \, V_{j_\nu}(\mu_\nu|z_\nu) \
\brangle^H_{(\lambda, \varpi,\tau)}  \ =\ {\cal  C } \,
|\Theta^g_N(u,y_i,z_\nu,\tau)|^2 \ \blangle \, \prod_{\nu=1}^N \,
V_{\alpha_\nu} (z_\nu) \ \prod_{i=1}^{N+2(g-1)} \,
V_{-1/2b}(y_i)\, \brangle^{L}_\tau
$$
where the pre-factor $\Theta_N$ takes the form
\begin{align} \label{Thetag}
 |\Theta^g_N(u,y_i,z_\nu,\tau)|^2
  &= ( |u |^2 e^{S_g})^{(1-g)} \prod_{i =1}^N H(z_i)
  ^{ - 1 - \frac{1}{b^2}}
   \prod_{k = 1}^{N+2(g-1)} H (y_k)^{1 + \frac{1}{b^2}}\,  \times
   \\[2mm]
  & \times \, \prod_{r < s}^N F ( z_r , z_s )^{\frac{1}{2 b^2}}
      \prod_{k < l}^{N+2(g-1)} F (y_k , y_l )^{\frac{1}{2 b^2}}
      \prod_{r=1}^N \prod_{k=1}^{N+2(g-1)}
       F (z_r , y_k)^{- \frac{1}{2 b^2}}\  ~, \nonumber
\end{align}
and we collected all the remaining terms in the quantity ${\cal
C}$,
$$ {\cal C}\  =\  e^{- S_{\rm sol} + (\frac{3}{2} + \frac{3}{4b^2}) U_g}\,
 |\det \partial |_{\lambda}^{-2}\ \ . $$
Before we conclude, let us observe that the constant ${\cal C}$
may  be written as the ratio between partition function $Z^L_0 =
Z^{\text{LD}}_{Q_\varphi}$ of a linear dilation with background
charge $Q_\varphi$ and the product $Z_0^H = |Z^{\beta\gamma}|^2
Z^{\text{LD}}_{Q_\phi}$ where $Z^{\beta\gamma}$ is the partition
function of a chiral $\beta\gamma$ system.  
In order to see this,
we recall that the partition function in both $H_3^+$ model and
Liouville field theory acquires it's leading (divergent)
contribution from the asymptotic region where the interaction is
negligible. 
Hence, we have
\begin{align}
Z^H_0 \ &= \  \int {\cal{D}} \phi_q {\cal{D}^\lambda} \c
 {\cal{D}^\lambda} \b \ e^{-
S_{\rm sol} - \frac{1}{2\pi} \int d^2w \, \left( \bpl \phi_q \pl
\phi_q -  \b \bpl \c - \bb \pl \bc + \frac{Q_\phi}{4} \, \sqrt{g}
{\cal R}\, \phi_q \right) } \ = 
 \nonumber \\[2mm]
&=\ |\det \partial |^{-2}_{\lambda} e^{-S_{\rm sol}}
  \int {\cal{D}} \phi_q
e^{ - \frac{1}{2\pi} \int d^2w \, \left( \bpl \phi_q \pl \phi_q +
\frac{Q_\phi}{4} \, \sqrt{g} {\cal R}\, \phi_q \right) } \  ~.
\end{align}
To further re-write the partition function $Z^H_0$, we must now
change the background charge from $Q_\phi$ to $Q_\varphi$. We can
achieve this using a result of \cite{Verlinde} on correlation
functions in a linear dilaton theory. When applied to the path
integral in the formula for $Z^H_0$, it formally reads
\begin{align}
  & \int {\cal{D}} \phi_q
e^{ - \frac{1}{2\pi} \int d^2w \, \left( \bpl_q \phi \pl \phi_q +
 \frac{Q_\phi}{4} \, \sqrt{g} {\cal R}\, \phi_q \right) }\ = \
\frac{e^{-\frac{3}{4} Q_\phi^2 U_g}}{|\det \partial|^2}\ = \
\nonumber \\[2mm]
& \ \ \ \ = \ e^{-\frac{3}{4} (Q^2_\phi-Q_\varphi^2) U_g}\int
{\cal{D}} \varphi e^{ - \frac{1}{2\pi} \int d^2w \, \left( \bpl
\varphi \pl \varphi +
 \frac{Q_\varphi}{4} \, \sqrt{g} {\cal R}\, \varphi \right) } \ = \
e^{\left( \frac{3}{2} + \frac{3}{4b^2} \right) U_g} \, Z^L_0
 \end{align}
where in the process of the calculation we inserted the explicit
expression for the background charges $Q_\phi$ and $Q_\varphi$.
Combining the previous two equations, we have shown,
\begin{equation} \label{partitionfct}
 Z^H_0 \ = \ |\det \partial |^{-2}_{\lambda} e^{-S_{\rm sol} +
\left( \frac{3}{2} + \frac{3}{4b^2} \right) U_g}\ Z^L_0 \ \ .
\end{equation}
In conclusion, our final result for the relation between
normalized correlation functions in the $H_3^+$ model and in
Liouville field theory reads \begin{equation} \label{corrga}
\frac{1}{Z^H_0}\, \blangle \ \prod_{\nu=1}^{N} \,
V_{j_\nu}(\mu_\nu|z_\nu) \ \brangle^H_{(\lambda, \varpi,\tau)}  \
= \ |\Theta^g_N(u,y_i,z_\nu,\tau)|^2 \ \frac{1}{Z^L_0}\, \blangle
\, \prod_{\nu=1}^N \, V_{\alpha_\nu} (z_\nu)\hspace*{-3mm}
\prod_{i=1}^{N+2(g-1)}\hspace*{-2mm} V_{-1/2b}(y_i)\,
\brangle^{L}_\tau\ \
\end{equation}
where the function $\Theta^g_N$ is given by eq.\ (\ref{Thetag}).
Let us also recall that the correlation functions of primaries
(\ref{vertex}) in the WZNW model at level $k = b^{-2} + 2$ depend
on the $g$ twists $\lambda_k$ and on $g-1$ zero modes $\varpi_l$
in addition to the surface moduli. On the Liouville side, we
compute the correlation functions of primaries (\ref{vertexL})
with a background charge $Q^L = Q_\varphi = b + 1/b$ and with bulk
cosmological constant $\mu_B = 4b^2$. The momenta $\alpha_\nu$ are
related to $j_\nu$ through eq.\ (\ref{ajrel}). The remaining
momenta $\mu_\nu$ in the WZNW model along with the $2g-1$ moduli
$\lambda_k$ and $\varpi_l$ determine the insertion points $y_i$ of
$N + 2(g-1)$ degenerate Liouville fields and a factor $u$ that we
absorbed in the definition of $\Theta_N^g$. Finally, we stress
that $Z^H_0$ and $Z^L_0$ are partition functions of free field
theories. They agree with those of the $H^+_3$ and Liouville
theory, respectively, if and only if we consider the theory on a
surface of genus $g=1$. For higher genus, our relation
(\ref{corrga}) shows that the partition function $Z^H$ of the
$H^+_3$ model is related to a $2g-2$ point function in Liouville
field theory.

\subsection{Free field theory computations}

In this subsection, we explain how to compute the residues of the
first order poles in WZNW correlation function from free field
theory (see also our discussion at the end of section 2). We then
determine the corresponding quantities for correlators in
Liouville field theory and show that the results agree with our
relation (\ref{corrga}) for the full correlators.%
\smallskip

As we have sketched in section 2.3 it is possible to compute the
residues of poles in the $H^+_3$ correlation functions by
inserting powers of the screening charges into correlators of a
linear dilaton $\phi$ and a $\beta\gamma$ system. For the rest of
this section we shall fix the world-sheet metric such that $\rho =
1$. The vertex operators and the usual screening charge take the
same form as above
\begin{align}
 V_j ( \mu | z ) &\equiv\  | \mu |^{ 2 j + 2}
  e^{ \mu \gamma - \bar \mu \bar \gamma} e^{2b ( j +1 ) \phi} ~,
   &S &= \ \int d^2 w S(w) \ = \ - \int d^2 w \beta \bar \beta
     e^{2 b \phi(w,\bar w)} ~.
\end{align}
The $N$-point correlation function has a pole at $\sum_\nu
(j_\nu + 1) = 1 - g - s$ $(s \in {\mathbb
Z}_{\geq 0})$, whose residue is obtained by integrating the following
correlators over the positions $w_k$
\begin{align}
 \blangle \prod_{\nu = 1}^N V_{j_\nu} (\mu_\nu | z_\nu )
         \prod_{k = 1}^{s} S (w_k) \brangle
 = \prod_{\nu = 1}^N | \mu_\nu |^{ 2 j_\nu + 2}
   \blangle \prod_{\nu = 1}^N e^{2 b ( j_\nu + 1) \phi ( z_\nu ,\bar z_\nu)}
         \prod_{k = 1}^{s} e^{2 b \phi (w_k , \bar w_k)}
         \brangle\ \times
  \nonumber  \\[2mm] \times\
   \blangle \prod_{\nu = 1}^N e^{\mu_\nu \gamma (z_\nu )}
         \prod_{k = 1}^{s} \beta (w_k) \brangle
   \blangle \prod_{\nu = 1}^N
     e^{ - \bar \mu_\nu \bar \gamma (\bar z_\nu )}
         \prod_{k = 1}^{s} [- \bar \beta (\bar w_k)] \brangle   ~.
         \label{H3+}
\end{align}
Throughout this entire subsection, correlation functions are
properly normalized such that the expectation value of the
identity is trivial rather than the partition function.
Since the free boson $\phi$ is subject to a background charge $Q =
Q_\phi = b$, the contribution from the linear dilaton theory can
be computed as
\begin{align}
  \blangle \prod_{\nu = 1}^N
   e^{2 b ( j_\nu + 1) \phi ( z_\nu ,\bar z_\nu)}
         \prod_{k = 1}^{s} e^{2 b \phi (w_k , \bar w_k)}   \brangle
  \ = \ \Lambda_{\text{sol}}\,
    \blangle \prod_{\nu = 1}^N
   e^{2 b ( j_\nu + 1) \phi_q ( z_\nu ,\bar z_\nu)}
         \prod_{k = 1}^{s} e^{2 b \phi_q (w_k , \bar w_k)}   \brangle
         ~,
\end{align}
where the twisted zero mode of $\phi$ contributes the factor
\begin{align} \label{Lambda}
 \Lambda_{\text{sol}} \ =\
  \prod_{\nu=1}^N
   e^{2 b ( j_\nu + 1) \phi_{\text{sol}} ( z_\nu ,\bar z_\nu)}
  \prod_{k = 1}^{s} e^{2 b \phi_{\text{sol}} (w_k , \bar w_k)}\ .
\end{align}
The correlation function of the single valued fluctuation
field $\phi_q = \phi - \phi_{\text{sol}}$ may be expressed through
the functions $F$ and $H$, see appendix \ref{riemann},
\begin{align}
   & \blangle \prod_{\nu = 1}^N
   e^{2 b ( j_\nu + 1) \phi_q ( z_\nu ,\bar z_\nu)}
         \prod_{k = 1}^{s} e^{2 b \phi_q (w_k , \bar w_k)}   \brangle
  \ = \ \prod_{\nu = 1}^ N H ( z_\nu)^{2 b^2 (j_\nu + 1 )}
    \prod_{k = 1}^s H (w_k)^{2 b^2}\ \times \label{ld} \\[2mm]
   & \ \ \ \ \ \times \
   \prod_{\nu < \mu}^N F(z_\nu . z_\mu)^{ - 2 b^2 (j_\nu +1)(j_\mu + 1)}
  \prod_{\nu=1}^N \prod_{k=1}^s F(z_\nu , w_k )^{ - 2 b^2 (j_\nu + 1)}
   \prod_{k < l}^s F ( w_k , w_l )^{ - 2 b^2}
    \ ~. \nonumber
\end{align}
The factors in the second line of eq.\ (\ref{H3+}) may be
evaluated using the same formulas we employed in our path integral
derivation. In particular, with the help of formula (\ref{svg1})
we can conclude that
\begin{eqnarray*}
 \blangle \prod_{\nu = 1}^N e^{\mu_\nu \gamma (z_\nu )}
         \prod_{k = 1}^{s} \beta (w_k) \brangle
  & = & \prod_{k=1}^s
  \left[ \sum_{\nu=1}^N \mu_\nu \sigma_{\lambda} ( z_\nu , w_k )
        + \sum_{\sigma=1}^{g-1} \varpi_\sigma \omega^\lambda_\sigma
   \right]  = \\[2mm]
  & = & \prod_{k=1}^s
   \left[ u \frac{ \prod_{i=1}^{N+ 2(g-1)}  E( y_i ,w_k )
     \sigma (w_k) ^2 }
        {\prod_{\nu=1}^N E (z_\nu , w_k ) }
       \right] \ .
\end{eqnarray*}
Utilizing the relation between the prime form $E$ and
the special function $F$ (see appendix \ref{riemann}) we conclude
\begin{align}
 & \blangle \prod_{\nu = 1}^N e^{\mu_\nu \gamma (z_\nu )}
         \prod_{k = 1}^{s} \beta (w_k) \brangle
 \blangle \prod_{\nu = 1}^N e^{- \bar \mu_\nu \bar \gamma (\bar z_\nu )}
         \prod_{k = 1}^{s} [ -\bar \beta (\bar w_k) ] \brangle =
  \nonumber \\[2mm]
 & \qquad \qquad = |u|^{2s} \prod_{k=1}^s \left[
   \frac{ \prod_{i=1}^{N+ 2(g-1)} | E( y_i ,w_k ) |^2
    | \sigma (w_k)|^4}
        {\prod_{\nu =1}^N |E (z_\nu , w_k )|^2 } \right] = 
  \label{equality}
 \\[2mm]
 &\qquad \qquad =
  |u|^{2s} \prod_{k=1}^s e^{ S_g - 2b \phi_{\rm sol}(w_k , \bar w_k)}
   \left[
   \frac{ \prod_{i=1}^{N+2(g-1)} F( y_i , w_k )
     H (w_k)^2}
        { \prod_{\nu=1}^N F (z_\nu , w_k ) }  \right]
  \nonumber
~.
\end{align}
Equations (\ref{Lambda}), (\ref{ld}) and (\ref{equality}) provide
all the information we need in order to determine the residue
(\ref{H3+}) of the WZNW correlator at $\sum_\nu (j_\nu + 1) = 1 -
g - s$.
\medskip

We would like to rewrite the correlation functions in $H^+_3$
model in terms of Liouville theory. The background charge of
Liouville theory is assumed to be $Q = b + 1/b$. We use the
following vertex operators and Liouville screening charge,
\begin{equation}
 V_{\alpha} (z , \bar z) = \
  e^{ 2 \alpha \varphi  (z , \bar z) } ~, \qquad
 S = \ \int d^2w S(w) \ = \ \int d ^2 w e^{2 b \varphi  (w , \bar w)}
\end{equation}
with $\alpha = b(j + 1) + 1/(2b)$. According to the general result
(\ref{corrga}), it should be possible to reproduce the residues
computed in the previous subsection from the expression
\begin{align}
 |\Theta^g_N|^2  &\ \blangle \prod_{\nu=1}^N
  e^{2 \alpha_\nu \varphi (z_\nu , \bar z_\nu ) }
  \prod_{i =1}^{N+2(g-1)} e^{- \frac{1}{b} \varphi (y_i , \bar y_i )}
  \prod_{k =1}^s e^{ 2 b \varphi (w_k , \bar w_k )}
 \brangle
\end{align}
where $\Theta_N^g$ is given in eq.\ (\ref{Thetag}) and the
correlation function is evaluated in a linear dilaton background
with background charge $Q = Q_\varphi = b + 1/b$. Using the same
formulas as in the previous subsection, we find
\begin{align}
 &\blangle \prod_{\nu=1}^N e^{2 \alpha_\nu \varphi (z_\nu , \bar z_\nu ) }
  \prod_{i =1}^{N+2(g-1)}
  e^{- \frac{1}{b} \varphi (y_i , \bar y_i )}
  \prod_{k =1}^s e^{ 2 b \varphi (w_k , \bar w_k )}
  \brangle =  \nonumber \\[2mm]
 & \qquad =  \prod_{\nu < \mu}^N F( z_\nu , z_\mu )^
  {- 2 ( b(j_\nu + 1) + \frac{1}{2b})( b (j_\mu + 1) + \frac{1}{2b})}
  \prod_{\nu=1}^N \prod_{i =1}^{N+2(g-1)}\!\!\!\!\!
 F( z_\nu , y_i )^{j_\nu +1 + \frac{1}{2 b^2}}
 \prod_{i < j}^{N + 2(g-1)} \!\!\!\!\!F ( y_i , y_ j )^{ - \frac{1}{2 b^2}}
  \times  \nonumber \\[2mm] & \qquad \times\,
 \prod_{\nu=1}^N \prod_{k=1}^s
 F ( z_\nu , w_k )^{- 2 b^2 ( j_\nu + 1) - 1}
  \prod_{i = 1}^{N+2(g-1)} \prod_{k=1}^s F ( y_i , w_k )
  \prod_{k < l}^s F ( w_k , w_l )^{- 2 b^2} \times
   \nonumber \\[2mm]  & \qquad \times\,
   \prod_{\nu =1}^N H(z_\nu)
  ^{ 2 ( b^2 + 1)(j_\nu + 1) + 1 + \frac{1}{b^2}}
   \prod_{i = 1}^{N+2(g-1)}\!\!\!H (y_i)^{- 1 - \frac{1}{b^2}}
   \prod_{k = 1}^s H (w_k)^{2 b^2 + 2} ~.\nonumber
\end{align}
Our aim now is to replace the factors $F(z_\nu,y_i)^{j_\nu+1}$
which do neither appear in $\Theta_N^g$ nor in the residues of the
WZNW model. Using the relation (\ref{muE}) with $\rho = 1$, we
find
\begin{align}
 \prod_{\nu=1}^N \prod_{k=1}^{N+2(g-1)} E ( z_\nu , y_k)^{j_\nu + 1}
  \sigma (z_\nu) ^ {2 j_\nu + 2} \ =\
  u^{s + g - 1} \prod_{\nu=1}^N \mu^{j_\nu +1}
  \prod_{\mu < \rho}^N E (z_\mu , z_{\rho} )^{j_\mu + j_{\rho} + 2}
 ~,
  \label{mu}
\end{align}
or, equivalently,
\begin{align}
& \prod_{\nu=1}^N \prod_{k=1}^{N+2(g-1)} F ( z_\nu , y_k)^{j_\nu +
1}
  H (z_\nu) ^ {2 j_\nu + 2} = \nonumber \\ &=
  (|u|^2 e^{S_g})^{s + g - 1}
  \prod_{\nu=1}^N |\mu|^{2j_\nu +2}
  e^{2 b(j_\nu + 2) \phi_{\rm sol} (z_\nu , \bar z_\nu)}
  \prod_{\mu < \rho}^N F (z_\mu , z_\rho )^{j_\mu + j_\rho + 2} \
  .
  \label{mu2}
\end{align}
In deriving this equation we have made use of the equality
$\sum_\nu (j_\nu + 1) = 1 - g - s$. Inserting \eqref{mu2} into
the linear dilaton correlator we rewrite the residues of
correlation functions in Liouville theory in an appropriate
form. Once these are multiplied with our function $\Theta_N^g$
these reproduce the results of the previous subsection and
thereby confirm nicely the outcome of our general path
integral derivation.

\section{Conclusions and Outlook}

In this work we proposed a new and elegant path integral 
derivation for the correspondence between local correlation 
functions in $H^+_3$ WZNW model and Liouville field theory. 
Our results reproduce the findings of \cite{RT} for 
correlators on the sphere and generalize them to surfaces 
of arbitrary genus $g$. Correlation functions of the $H^+_3$ 
WZNW model are determined from Liouville field theory through 
eq.\ \eqref{corrga}. Physical correlators are obtained in the 
limit $\lambda_k \rightarrow 0$ of vanishing twist parameters 
and after integration over $\varpi_\sigma$. A correspondence
between residues of correlation functions is encoded in eq.\ 
\eqref{corrga} and it was verified explicitly through free 
field calculations. For the torus, we have also explained 
how to map the BPZ equations for Liouville theory to the KZB 
equations in WZNW models. The extension of this analysis to 
higher genus was not addressed. We believe that an explicit 
comparison along the lines of section 3.2 is possible, though 
rather cumbersome. 
\smallskip 

It seems appropriate to add a few comments on our $2g-1$ moduli 
$\lambda_k, \varpi_\sigma$ and to explain their relation with 
the twist parameters introduced by Bernard \cite{Bernald2}. We 
recall that Bernard's constructions involve a $3g$-dimensional 
space of twists. It parametrizes a set of $g$ group elements 
which determine boundary conditions of currents along the 
$\beta$-cycles. Our coordinates $\lambda_k$ correspond to 
very special twists with group elements of the form $g_k 
\sim \exp(- 2 \pi i \lambda_k J^0)$ where $J^0$ is the 
Cartan generator. Consequently, insertions of the
current $J^0(w)$ into correlators of the WZNW model may be
converted into the action of some differential operator. The
latter has exactly the same form as in Bernard's work (see 
eqs.\ (4.16) and (4.17) of \cite{Bernald2}). For insertions 
of the component $J^-(w)$, the story is a bit different. In 
this case, we may use the relation $J^-(w) =\beta(w)$ along 
with our equation (\ref{bspbg}) to derive
\begin{eqnarray} \label{Jp}
\left\langle \, J^-(w)\, \prod_{\nu=1}^N \,
V_{j_\nu}(\mu_\nu|z_\nu)
 \right\rangle^{H}_{(\lambda,\varpi,\tau)}&= & \left(
\sum_{\nu=1}^N \mu_\nu \sigma_{\lambda} (w , z_\nu) +
\sum_{\sigma=1}^{g-1} \varpi_\sigma \omega^\lambda_\sigma (w) \right)
\ \Omega^H_{(\lambda,\varpi,\tau)} \\[2mm]
\mbox{where } \ \ & & \Omega^H_{(\lambda,\varpi,\tau)} \, = \,
\left\langle \, \prod_{\nu=1}^N \, V_{j_\nu}(\mu_\nu|z_\nu)
 \right\rangle^{H}_{(\lambda,\varpi,\tau)} ~.
\end{eqnarray}
On the right hand side, the complex numbers $\varpi_\sigma$ that
multiply the holomorphic one-forms appear in place of
differentiation with respect to twist parameters in Bernard's
work. In this sense, our $\varpi_\sigma$ are dual twist
parameters. We are not convinced that insertions of the third
current $J^+(w)$ can similarly be replaced by the action of some
operator. Let us point out, however, that this is not crucial for
a successful match between the BPZ and KZB-type differential
equations. In fact, only certain combinations of the KZB equations
appear in this context. The relevant ones emerge from inserting
the Sugawara tensor at the points $y_i$ at which $J^-$ vanishes.
Hence, the term $J^+ J^-$ drops out. 
\smallskip

As explained e.g.\ in \cite{Frenkel} (see also references therein), 
one of the ramifications of the geometric Langlands program involves 
conformal blocks of WZNW models at the so-called critical level and 
their relation with certain classical $\mathcal{W}$ algebras. In the case 
of the $H^+_3$ model, the critical level is $k=2$. Hence, we reach 
this point in the limit $b \rightarrow \infty$ in which the associated 
Liouville theory becomes classical. For genus $g=0$ correlation 
functions in the $H^+_3$-Liouville correspondence, the limit of 
infinite parameter $b$ was analyzed in detail in \cite{RT}. It 
might be rewarding to carry out a similar investigation for 
surfaces of higher genus $g \neq 0$. In this case, the Gaudin
Hamiltonians that emerge from the critical WZNW model on the 
sphere get replaced by Hamiltonians of Hitchin's integrable 
system. 
\smallskip 

The relation (\ref{corrga}) between correlation functions in the 
$H^+_3$ model and Liouville field theory may be regarded as an  
`off-critical' (and non-chiral) version of the geometric Langlands 
program. Let us recall that conformal blocks of the WZNW model 
diagonalize the action of the current algebra on the fusion 
product of its representation spaces in the same sense in which 
Clebsch-Gordan maps (block-) diagonalize the action of a Lie 
algebra on tensor products. When we are dealing with Lie algebras, 
the geometric Langlands program achieves more: it provides a 
distinguished basis in the tensor product consisting 
of eigenvectors of a classical $\mathcal{W}$-algebra. Once we 
go off-critical, the classical $\mathcal{W}$-algebra becomes 
quantum. In the case considered here, the $\mathcal{W}$-algebra 
is the Virasoro algebra. As usual, the action of the 
$\mathcal{W}$-algebra is block diagonalized by its conformal 
blocks. Putting all this together, an off-critical version of the 
geometric Langlands program should single out a distinguished basis 
for WZNW conformal blocks which may be expressed directly through 
conformal blocks of the $\mathcal{W}$-algebra. Our main result 
eq.\ (\ref{corrga}) claims that for the $H^+_3$ model such a 
basis is given by the WZNW correlators on the left hand side.
Let us stress that the proper basis is found for (twisted) 
correlators on any closed Riemann surface. It seems likely 
that a similar off-critical version of the geometric 
Langlands correspondence exists for other Lie algebras 
(see also comments below). 
\smallskip

There are several extensions of our results that seem worthwhile 
being analyzed. To begin with, it would be interesting to study 
correlation functions on surfaces with boundaries. For the WZNW model,
maximally symmetric boundary conditions were found in \cite{PST}.
Using new boundary theories for $\beta \gamma$ systems 
(see \cite{CQS1}) along with ideas from a forthcoming paper on branes 
in the GL(1$|$1) model \cite{CQS2}, a first order formulation  
for correlators with insertions of both bulk and boundary operators
can be developed. An evaluation along the lines we presented above
should then relate these to correlation functions in boundary
Liouville theory \cite{FZZ,TLb,ZZ}. For some disc amplitudes,
such relations between correlators on a surface with boundary
have been proposed in \cite{HR}.
\smallskip

More importantly, it is very tempting to address generalizations
to WZNW models of rank $r > 1$. First order formulations for
models with higher rank are certainly known (see e.g. 
\cite{Petersen:1997ui} and references therein) and
it is likely that these may be employed to reduce correlators of
WZNW primaries to correlation functions in conformal Toda
theories. The evaluation of the corresponding WZNW path integral,
however, requires significant new ideas, mainly because the
nilpotent part of higher rank algebras is no longer abelian. This
is directly linked to a non-linear dependence of the Kac-Wakimoto
like action functionals on some of the fields $\gamma$. We plan to
return to these issues in the near future.


\acknowledgments

 We are grateful to J\"org
Teschner and Sylvain Ribault for many useful discussions and
comments. This work of YH was supported by an JSPS Postdoctoral
Fellowship for Research Abroad under contract number H18-143.

\appendix

\section{Differential equations for genus one case}
\label{apg1}

In this appendix we would like to demonstrate that the
differential operators (\ref{BPZop}) and (\ref{KZBop}) are related
through equation \eqref{diffg1}. As we described in the main text,
it is important to first replace the derivatives $\partial^H_\nu =
\partial/\partial z_\nu$ which are evaluated for fixed $\mu_\nu$
in terms of $\partial_\nu^L$. This is achieved with the help of
formula (\ref{deltai}). In order to verify the latter, we must
show that
\begin{equation}
 \delta_i \mu_\nu =\ 0 ~, \qquad
   \delta_i \lambda =\ 0 ~.
 \label{deltaa}
\end{equation}
The second equation can be shown trivially with eq.\ \eqref{rel1}.
The first property of $\delta_i$ may be established as follows.
With eq.\ \eqref{rel1} we find
\begin{align}
 \mu_\nu^{-1} \delta_i \mu_\nu
 \ = \ \xi (y_i , z_\nu ) ( \sum_j \xi (z_\nu,y_j)
                      - \sum_{\mu \neq \nu} \xi (z_\nu ,z_\mu ) ) +
 \sum_{\mu \neq \nu} \xi (y_i , z_\mu)  \xi (z_\nu,z_\mu) - \label{mui}\\
   - \sum_{\mu} \xi (y_i , z_\mu) \xi (z_\nu , y_i )
 + \sum_{j \neq i} \xi (y_i , y_j) \left(  \xi (z_\nu,y_i)
                      -  \xi (z_\nu , y_j ) \right) ~.
 \nonumber
\end{align}
The right hand side is a double periodic function of $y_i$, which
could become singular at $y_i \sim y_j,z_\nu$. We can analyze the
singular behavior
of $\mu_\nu^{-1} \delta_i \mu_\nu$ with the help of the following
expansions,
\begin{align}
\xi (z,z') &=\ \frac{1}{z - z'} - 2 (z-z') \eta_1 + {\cal
O}((z-z')^2) ~,
 \nonumber
\\[2mm]
\partial_z \xi (z,z') &=\ - \frac{1}{(z-z')^2}
 - 2 \eta_1 + {\cal O} (z-z')~,
 \label{xib}
 \\[2mm]
\xi (z,z') ^2  &=\ \frac{1}{(z-z')^2}
 - 4 \eta_1 + {\cal O} (z-z')
 \nonumber
\end{align}
at $z \sim z'$. In fact, using the above expansions for $\xi$ one
can show that  $\mu_\nu^{-1} \delta_i \mu_\nu$ has no
singularities when two of the insertion points $y_i$ and
$z_\nu$ approach each other. Since
the whole expression is double periodic and free of singularities,
the function should be constant, independent of $y_i,z_\nu$.
Therefore, it suffices to calculate it at one single point.
Let us set $y_i = z_\nu$, then we find
\begin{align}
 \mu_\nu^{-1} \delta_i \mu_\nu
 \ = \ \sum_{\mu \neq \nu}
 \frac{\theta ^{''} (z_\nu - z_\mu)}{\theta (z_\nu -z_\mu)}
 - \sum_{j \neq i} \frac{\theta ^{''} (z_\nu - y_j)}{\theta (z_\nu -y_j)} ~.
\end{align}
Here, we have used the following expansion around $t \sim z_\nu$,
$$
\frac{\theta ' (t - z_\mu)}{\theta (t -z_\mu)} \ = \ \frac{\theta
' (z_\nu - z_\mu)}{\theta (z_\nu -z_\mu)}
 + \left[ \frac{\theta ^{''} (z_\nu - z_\mu)}{\theta (z_\nu -z_\mu)}
 - \left( \frac{\theta ' (z_\nu - z_\mu)}{\theta (z_\nu -z_\mu)} \right)^2
  \right] (t - z_\nu) + {\cal O} ( (t - z_\nu)^2 ) ~.
$$
Furthermore, we can set each of $y_j$ $(j \neq i)$ to be one of
$z_\mu$ $(\mu \neq \nu)$ because the above quantity does not
depend on $y_j$ either. {}From the above equation we can see that
the equation $ \mu_\nu^{-1} \delta_i \mu_\nu = 0$ is indeed
satisfied.
\smallskip

There appears another derivative in the differential operator
$D^{\KZB}_i$ for torus correlation functions, namely the
derivative $\partial_\tau$ with respect to the modulus. We would
like to check that it acquires no corrections when we switch from
the variables $\mu_\nu$ to $y_i,u$, or, in other words,
\begin{align}
 \frac{\partial}{\partial \tau} \mu_\nu\ = \ 0 ~.
\end{align}
In order to see this, we compute
\begin{align}
  4 \pi i  \mu_\nu^{-1}  \frac{\partial}{\partial \tau} \mu_\nu
  \ =\ \sum_i \frac{\theta ^{''} (z_\nu - y_i)}{\theta (z_\nu - y_i)}
    + 6 \eta_1
 - \sum_{\mu \neq \nu}
 \frac{\theta ^{''} (z_\nu - z_\mu)}{\theta (z_\nu - z_\mu)} ~.
\end{align}
Here we should notice that
\begin{align}
 4 \pi i \frac{ \partial_\tau \theta (z) }{\theta (z) }
 &= \frac{ \theta^{''} (z) }{\theta (z) }
 = \partial_z {\xi (z,0) } + \xi (z,0)^2 ~,
 &4 \pi i \frac{ \partial_\tau \theta (z) }{\theta (z) }
 &= - 6 \eta_1 + {\cal O} (z) ~.
\end{align}
{}These formulas show that $\mu_\nu^{-1} \partial_\tau \mu_\nu$
has no singularities. As before we may evaluate the derivative at
e.g.\ $y_i = z_i$ and then find indeed ${\mu_\nu}^{-1}
\partial_\tau \mu_\nu = 0$.
\smallskip

In order to show \eqref{diffg1}, we have to understand the
properties of the function $\Theta_N '$ that we defined in eq.\
\eqref{twistg1},
\begin{align}
  &\Theta '_N \ =\   \theta ' (0) ^{\frac{N}{2 b^2}}
      \prod_{\nu < \mu}^N  \theta ( z_\nu - z_\mu ) ^{\frac{1}{2 b^2}}
      \prod_{i < j}^{N}  \theta (y_i - y_j ) ^{\frac{1}{2 b^2}}
      \prod_{\nu,i=1}^N  \theta (z_\nu - y_i) ^{- \frac{1}{2 b^2}} ~.
\end{align}
Conjugation of the various derivatives by the factor $\Theta_N'$
gives
\begin{align*}
 &{\Theta '_N}^{-1} \frac{\partial}{\partial z_\nu} \Theta '_N
 \ = \ \frac{\partial}{\partial z_\nu}
 + \frac{1}{2 b^2 } \sum_{\mu \neq \nu}  \xi (z_\nu , z_\mu )
 - \frac{1}{2 b^2 } \sum_{i}  \xi (z_\nu , y_i ) ~, \\[2mm]
 &{\Theta '_N}^{-1} \frac{\partial}{\partial y_i} \Theta '_N
 \ =\  \frac{\partial}{\partial y_i}
 + \frac{1}{2 b^2 } \sum_{j \neq i}  \xi (y_i , y_j )
 - \frac{1}{2 b^2 } \sum_{\nu}  \xi (y_i , z_\nu ) ~,\\[2mm]
 &{\Theta '_N}^{-1} \frac{\partial}{\partial \tau} \Theta '_N
 \ = \ \frac{\partial}{\partial \tau} + T ~, \\[4mm]
 &T = \ \frac{N}{2 b^2} \frac{\partial_\tau \theta ' (0)}{\theta ' (0)}
 + \frac{1}{2 b^2 } \sum_{\nu < \mu }
   \frac{\partial_\tau \theta (z_\nu - z_\mu)}{\theta (z_\nu - z_\mu)}
 + \frac{1}{2 b^2 } \sum_{i < j}
   \frac{\partial_\tau \theta (y_i - y_j)}{\theta (y_i - y_j)}
 - \frac{1}{2 b^2 } \sum_{\nu , i}
   \frac{\partial_\tau \theta (z_\nu - y_i)}{\theta (z_\nu - y_i)} ~.
\end{align*}
Moreover, we note that
\begin{align}
 &{\Theta '_N}^{-1} b^2 \frac{\partial ^2 }{\partial y_i ^2} \Theta '_N
 \ = \ b^2 \frac{\partial ^2}{\partial y_i ^2 }
 + \left( \sum_{j \neq i}  \xi (y_i , y_j )
 -  \sum_{\nu}  \xi (y_i , z_\nu ) \right)
\frac{\partial }{\partial y_i} +  \nonumber \\[2mm]
  &+ \frac{1}{2} \left(
   \sum_{j \neq i} \partial_i \xi (y_i , y_j )
   -  \sum_{\nu} \partial_i \xi (y_i , z_\nu )  \right)
 + \frac{1}{ 4 b^2 }\left(\sum_{j \neq i}  \xi (y_i , y_j )
 - \sum_{\nu}  \xi (y_i , z_\nu ) \right)^2  ~,
\end{align}
and
\begin{align}
 {\Theta '_N}^{-1} \delta_i \Theta '_N =
 &=\ \sum_{\nu} \xi (y_i , z_\nu)
 \left( \frac{\partial}{\partial z_\nu}
 + \frac{1}{2 b^2 } \sum_{\mu \neq \nu}  \xi (z_\nu , z_\mu )
 - \frac{1}{2 b^2 } \sum_{j}  \xi (z_\nu , y_j ) \right) +
 \nonumber \\[2mm]
 &+ \sum_{\nu} \xi (y_i , z_\nu)
 \left(  \frac{\partial}{\partial y_i}
 + \frac{1}{2 b^2 } \sum_{j \neq i}  \xi (y_i , y_j )
 - \frac{1}{2 b^2 } \sum_{\mu}  \xi (y_i , z_\mu )\right) -
 \nonumber \\[2mm]
 &-  \sum_{j \neq i} \xi (y_i , y_j)
 \left( \frac{\partial}{\partial y_i}
 + \frac{1}{2 b^2 } \sum_{k \neq i}  \xi (y_i , y_k )
 - \frac{1}{2 b^2 } \sum_{\nu}  \xi (y_i , z_\nu ) \right) +
 \nonumber \\[2mm]
 &+ \sum_{j \neq i} \xi (y_i , y_j)
 \left(  \frac{\partial}{\partial y_j}
  + \frac{1}{2 b^2 } \sum_{k \neq j}  \xi (y_j , y_k )
 - \frac{1}{2 b^2 } \sum_{\nu}  \xi (y_j , z_\nu )\right) ~.
 \nonumber
\end{align}
This long list of equations puts us into the position to finally
prove eq.\ \eqref{diffg1},
%
\begin{align}
 (\Theta ' _N \eta)^{-1} D_i^{\KZB} (\Theta ' _N \eta)
 - D_i^{\BPZ}
  \ = &- \frac{1}{4 b^2} (
  \sum_{j \neq i} \xi (y_i , y_ j ) - \sum_\nu \xi ( y_i , z_\nu ) )^2 +
  \label{KZBBPZ} \\[2mm]
  &+ \frac{1}{2 b^2 }\sum_\nu \xi (y_i , z_\nu )
   ( \sum_{\mu \neq \nu} \xi (z_\nu , z_\mu)  - \sum_j \xi (z_\nu , y_j ) ) +
\nonumber \\[2mm]
 &+ \frac{1}{2 b^2 }\sum_{j \neq i} \xi (y_i , y_j )
   ( \sum_{k \neq j} \xi (y_j , y_k) - \sum_\nu \xi (y_j , z_\nu ) ) +
\nonumber \\[2mm]
 & + \frac{3}{2b^2} \eta_1
 - \frac{3}{4b^2} \sum_{j \neq i } \partial_i \xi (y_i , y_j )
 + \frac{1}{4b^2} \sum_{\nu} \partial_i \xi (y_i , z_\nu )
 + T ~. \nonumber
\end{align}
There could be double or single poles at $y_i = z_\nu,y_j$, but we
can check that such terms are absent. Moreover, even if we regard
the expression \eqref{KZBBPZ} as a function of $y_j$ $(j \neq i)$,
there are no singular terms. Therefore, the problem is whether the
constant independent of $y_i,y_j$ vanishes or not. We set $y_j =
z_j$ for $j \neq i$ so that \eqref{KZBBPZ} becomes
\begin{align}
 &(\Theta '_N \eta )^{-1} D_i^{\KZB} ( \Theta_N ' \eta )
 - D_i^{\BPZ} = \nonumber \\[2mm]
  &= \frac{1}{4 b^2} \left(  \zeta (y_i , z_i )^2
 + \partial_{i} \zeta (y_i , z_i ) \right)
 + \frac{3}{2 b^2} \eta_1 -
   \frac{1}{2 b^2} \left( 6 \eta_1
    + \frac{\theta ^{''} (y_i - z_i )}{\theta (y_i - z_i) }\right)
    ~,
\end{align}
where the last term of the right hand side comes from $T$ and in
the derivation we have also used
$$ 2 \pi i  \eta (\tau )^{-1} \frac{\partial}{\partial \tau}
\eta (\tau) \ = \ - \eta_1 \ .
$$ Taking the limit
of $y_i \to z_i$, we indeed obtain the desired equation
(\ref{diffg1}).

\section{Theta functions on a general Riemann surface}
\label{riemann}

In this appendix we summarize some basic results concerning
a free boson on a Riemann surface of genus $g$. Basically,
our exposition follows the discussions in \cite{Verlinde}.
See also \cite{AMV,Fay,Mumford}.

\subsection{The prime form}

We consider correlation functions on a compact Riemann surface
$\Sigma$ of genus $g$ with a complex structure. Let us choose a
canonical basis of homology cycles $\alpha_k, \beta_k$
$(k=1,\cdots,g)$ satisfying
\begin{equation}
\oint_{\alpha_k} \omega_l =\  \delta_{kl} ~, \qquad \oint_{\beta_k}
\omega_l = \ \tau_{kl} ~,
\end{equation}
where $\omega_l (l=1,\cdots,g)$ denote the holomorphic one-forms
on $\Sigma$. The complex symmetric matrix $\tau_{kl}$ $({\rm Im}
\tau_{kl} > 0)$ is known as the period matrix. We now fix an
arbitrary point $p_0$ in $\Sigma$ and construct a map from the
universal cover $\tilde \Sigma$ of the surface $\Sigma$ to
$\mathbb{C}^g$,
\begin{align}
 z_k (p) \ = \ \int^{p}_{p_0} \omega_k
\end{align}
with $p,p_0$ on $\tilde \Sigma$. This embedding of $\tilde \Sigma$
into $\mathbb{C}^g$ is known as the Abel map. Functions on the
image of the Abel map descend to the surface $\Sigma$ if they are
periodic under all shifts of the form $z'_k = z_k + m_k +
\tau_{kl}n^l$ with integer coefficients $m_k, n^l$. In order to
construct a few basic objects on the surface $\Sigma$ and its
cover $\tilde \Sigma$, we recall the following definition of
Riemann's theta function,
\begin{align}
 \theta_\delta (z | \tau )
 \  = \ \sum_{n \in {\mathbb{Z}}^g}
  \, \exp i \pi [ (n + \delta_1 )^k \tau_{kl} (n + \delta_1 )^l
   + 2 (n+\delta_1 )^k ( z + \delta_2 )_k ]\  ~,
\end{align}
where $\delta_k = (\delta_{1k} , \delta_{2k} )$ with $\delta_{1k},
\delta_{2k} = 0, 1/2$ denotes the so-called {\em spin structure}
along the $\alpha_k$ and $\beta_k$ cycles. Under shifts along the
$2g$ fundamental cycles, $\theta_\delta$ behaves as
\begin{align}
  \theta_\delta (z + \tau n + m | \tau )
 \  = \ \exp [ - i \pi ( n^k \tau_{kl} n^l + 2 n^k z_k ) ]
    \exp [ 2 \pi i ( \delta_1^k m_k - \delta_2^k n_k ) ]
\theta_\delta (z | \tau )\  ~.
\end{align}
The Riemann vanishing theorem asserts that $\theta (z ,\tau )$
vanishes in a point $z$ on $\Sigma$ 
if and only if there
exists $g-1$ points $p_i$ on $\Sigma$ such that $z$ can be written in the
form
\begin{align}
 z \ = \ \Delta - \sum_{k=1}^{g-1} p_k \ ~,
\end{align}
where $\Delta$ is a fixed divisor on the surface $\Sigma$ that is
known as Riemann class. The right hand side of this equation could be
considered as an element of $\mathbb{C}^g$ through application of
the Abel map. Let us now introduce the following holomorphic
1/2-differential $h_\delta$
\begin{align}
 ( h_\delta (z) )^2 \ =\  \sum_k \, \partial_k \theta_{\delta} (0 | \tau)
  \omega_k (z)\,  ~.
\end{align}
$h_\delta$ is the essential building block for the important prime form
$E$
\begin{align} \label{prime}
 E (z , w )\  = \  \frac{\theta_\delta ( \int^z_w \omega | \tau)}
                   {h_\delta (z) h_\delta (w)}
\end{align}
which is defined for any odd spin structure $\delta$. The prime
form $E$ has weight $(-1/2 ,0) \times (-1/2 , 0)$ and near its
unique zero at $z = w$ one finds $E(z,w) \sim z - w$. Moreover,
$E$ is periodic under shifts $z_l$ along the $\alpha_k$-cycle
as $z'_l = z_l + n_l$ with $n_l = \delta_{l,k}$. On the
other hand, a non-trivial phase appears if we shift $z_l$ along the
$\beta_k$-cycle as $z'_l = z_l + \tau_{lk}$,
\begin{align}
 E ( z  + \tau_k , w )\  =\  -
\exp \left( - i \pi \tau_{kk}-2\pi \int^w_z \omega_k \right)
E(z,w)\  ~. \label{prime_factor}
\end{align}
On the left hand side of this equation, the objects $\tau_k$
denotes the $k^{th}$ column $\tau_{\cdot,k}$ of the period matrix
$\tau_{lk}$.

\subsection{Free linear dilaton theory}

Let us employ the prime form and some close relatives thereof to
spell out the $N$-point functions of a free bosonic field with
background charge $Q$. For the fluctuation around the zero mode,
the correlation functions can be given as \cite{Verlinde}
\begin{align}
 \blangle\,  \prod_{i=1}^N e^{2 \alpha_i \varphi} (z_i)
 \, \brangle \ = \
\prod_{i < j}\,  F(z_i , z_j )^{- 2 \alpha_i \alpha_j}\  \prod_{i}
\, H (z_i)^{2 Q \alpha_i}\  ~,
\end{align}
where we assume that
\begin{align}
 \sum_i \alpha_i \ = \ Q ( 1 - g )\  ~.
\end{align}
We have defined $F$ and $H$ as
\begin{align}
 F ( z , w ) \ = \ \exp
  \left( - 2 \pi {\rm Im} \int^z_w \omega^k\
 ({\rm Im} \tau )^{-1}_{kl}\  {\rm Im} \int^z_w \omega^l \right)
  | E (z , w ) |^2 \  ~,
\end{align}
and
\begin{align}
 H (z)\  =\  | \rho (z) |
 \exp \left( \frac{1}{16 \pi} \int d^2 w
 \sqrt{g} \mathcal{R} (w) \ln ( F(z,w))
 \right) \ ~.
\end{align}
Integration over the $g$ holomorphic forms $\omega_k$ furnishes an
element of $\mathbb{C}^g$ that can be multiplied with  $({\rm Im}
\tau )^{-1}$. As in \cite{Verlinde} we can rewrite the function
$H$ in a form
\begin{align}
 H ( z )\  = \ \exp
  \left( \frac{2 \pi}{g-1} {\rm Im} \int^\Delta_{(g-1)z}
 \omega^k\
 ({\rm Im} \tau )^{-1}_{kl} {\rm Im} \ \int^\Delta_{(g-1)z}  \omega^l \right)
  | \sigma (z  ) |^2\
  \label{hz}
\end{align}
which involves the Riemann class $\Delta$ that was introduced in
the previous subsection. Let now $p_k$ denote $g$ arbitrary points
on $\Sigma$. Then the function $\sigma (z)$ satisfies
\begin{align}
 \frac{\sigma (z)}{\sigma (w)}
 \  =\  \frac{\theta_0 ( z - \sum p_k + \Delta ) }
 {\theta_0 ( w - \sum p_k + \Delta ) } \prod_k
  \frac{E ( w , p_k )}{E ( z , p_k )}\  ~.
\end{align}
It is important to mention that $\sigma (z)$ is a
$g/2$-differential and that it has no zeros and poles. When
translated with the $k^{th}$ column vector of the period matrix,
$\sigma$ satisfies
\begin{align}
 \sigma ( z + \tau_k, w ) \ =\
\exp \left(  \pi i (g-1) \tau_{kk}  - 2 \pi i \int^\Delta_{ (g-1)
z } \omega_k  \right) \sigma (z,w) \ ~.
  \label{sigma_factor}
\end{align}
As before, we combined the matrix elements $\tau_{lk}$ for $l=1,
\dots,g$ into an element $\tau_k \in \mathbb{C}^g$.

\end{document}